\documentclass[aps,pra,twocolumn, amsmath, amssymb,superscriptaddress]{revtex4}
\usepackage{graphicx}
\usepackage{dcolumn}
\usepackage{bm}
\usepackage{indentfirst}

\begin{document}

\title{Dissipative pure-optical trap for laser cooling and trapping of neutral atoms}

\author{O.N. Prudnikov}
\email{oleg.nsu@gmail.com}
\affiliation{Institute of Laser Physics, 630090, Novosibirsk, Russia}
\affiliation{Novosibirsk State University, 630090, Novosibirsk, Russia}

\author{R. Ya. Ilenkov}
\affiliation{Institute of Laser Physics, 630090, Novosibirsk, Russia}
\affiliation{Novosibirsk State University, 630090, Novosibirsk, Russia}


\author{A. V. Taichenachev}
\affiliation{Institute of Laser Physics, 630090, Novosibirsk, Russia}
\affiliation{Novosibirsk State University, 630090, Novosibirsk, Russia}

\author{V. I. Yudin}
\affiliation{Institute of Laser Physics, 630090, Novosibirsk, Russia}
\affiliation{Novosibirsk State University, 630090, Novosibirsk, Russia}
\affiliation{Novosibirsk State Technical University, 630073, Novosibirsk,
Russia}

\author{S. N. Bagaev}
\affiliation{Institute of Laser Physics, 630090, Novosibirsk, Russia}

\date{\today}

\begin{abstract}
We show the possibility of implementing a deep dissipative optical lattice for neutral atoms with a macroscopic period. The depth of the lattice can reach magnitudes comparable to the depth of the magneto-optical traps (MOT), while the presence of dissipative friction forces allows for trapping and cooling of atoms.
The area of localization of trapped atoms reaches sub-millimeter size, and the number of atoms is comparable to the number trapped in MOT. As an example, we study lithium atoms for which the macroscopic period of the lattice $\Lambda=1.5$ cm. Such deep optical lattices with a macroscopic period open up possibility for developing effective methods for cooling and trapping neutral atoms without use of magnetic field as an alternative to MOT. This is important for developing compact systems based on cold atoms.
\end{abstract}


\maketitle


\section{Introduction}
Magneto-optical traps (MOT) are the basic instrument for laser cooling and trapping of neutral atoms \cite{raab1987,metcalf}. The operation of MOT is based on a combination of dissipative forces of spontaneous light pressure and a deep macroscopic potential (of the order of several $K$)  in a spatially non-uniform magnetic field. Cold atoms obtained in MOT have a wide range of applications, including the development of quantum sensors based on matter wave interference \cite{berman}, in laser metrology for developments  of ultra-precise frequency standards (atomic clocks) \cite{Katori2020,Lion2017,Ludlow2018,Nicholson} and others.
However, in many cases fast turn-off of magnetic field used in MOT and precise control over the residual magnetic field are required, which may be technically difficult to realize.  Therefore, the development of alternative methods of primary laser cooling and trapping of neutral atoms without using a magnetic field is an important direction in creating compact and mobile ultra-precision devices based on cold atoms.

It is well known that optical lattices allow for atom trapping without a magnetic field. Dissipative and non-dissipative optical lattices are distinguished \cite{metcalf,jessen96,Grynberg}. The regime of dissipative lattices \cite{metcalf,Grynberg,dal1989,dal1993EPL,pru2007j,pru2004,Dal85d,pru2007,pru2011}, which combines atom trapping and laser cooling, is realized for small detunings, when the light field frequency is close to the atom transition resonance with natural linewidth $\gamma$. However, since the frequency is chosen close to the atomic transition and the intensities of light waves required for deep cooling should be small, the depth of the dissipative lattice is usually small and comparable to the temperature of Doppler or sub-Doppler cooling. The absence of a deep macroscopic potential does not allow the use of such lattices for efficient primary laser cooling and trapping. Non-dissipative deep optical lattices are created by  intense light waves with a sufficiently large detuning. Their depth is determined by the technical capabilities of laser systems and can reach hundreds of microkelvin \cite{metcalf,Rassel2015,Katori2020}. However, dissipative laser cooling mechanisms are negligible here, which requires the use of MOT to load atoms into them.

The use of bichromatic waves opens up new possibilities to form a deep dissipative potential of macroscopic scale. The first theoretical studies of atomic kinetics in the presence of two monochromatic fields have been done in \cite{kaz87,Voitsekhovich88}, where the effect of rectification of dipole force was demonstrated. In such fields the atom acquires momentum $\Delta p = 2 \hbar k$ due to induced absorption of photons of one of the counter-propagating waves and induced emission into the opposite wave. Since the rate of induced processes is not directly related to the line width $\gamma$, the force on the atom can significantly exceed the spontaneous light pressure force from a single wave, which can be used for effective control of atomic beams \cite{ovch90,Soding97,Metcalf_He,Kitching2012}.

Moreover, in bichromatic field, the force averaged over the wavelength is not equal to zero, making it possible to create a deep optical potential with a macroscopic period $\Lambda = \pi/\Delta k$, determined by the spatial beating of the two frequency components ($\Delta k$ is difference in wave vectors). In this case, the parameters of the light field can be chosen so that the dissipative mechanisms of laser cooling lead to cooling and trapping of the atoms in the area of minimum of macroscopic potential \cite{pru2013,pru2017}. However, because both field frequency  are resonant with the same optical transition, the difference in wave vectors is extremely small, and $\Lambda$ is very large. For example, for a frequency difference of 10-100 MHz is required for near-resonant interaction the macroscopic period is about 1-10 meters. In such fields, the curvature in the minimum of the macroscopic potential  ($\propto \Delta k/k$) is small and does not allow to reach a distinct localization of atoms inside cell of centimeter size, since the spatial phase difference of the fields at such scales is practically unchanged.

In this work, to reduce the period of the macroscopic potential in a bichromatic field and increase its curvature, we propose to use light  waves, which are resonant to different transitions between fine and hyperfine components of atomic levels.
As an example, we investigate lithium atoms, for which the fine splitting of the $^2P_{3/2}$ and $^2P_{1/2}$ levels is about 10 GHz, determining the macroscopic lattice period of $\Lambda = 1.5$ cm. In this case, each field component is near-resonant to the $D_1$ and $D_2$ lines, which leads to significant dipole forces and allows for the realization of a deep optical potential with a macroscopic period even in low-intensity field.  The presence of dissipative effects in such an optical lattice leads to laser cooling and trapping of neutral atoms with a sub-millimeter localization area near the minimum of the macroscopic potential. The results of our theoretical analysis show that the depth of the macroscopic potential is comparable to the depth of the magneto-optical trap (approximately 1 K in temperature units or more), and laser cooling temperatures can reach sub-Doppler values. These studies open up the possibility of implementing deep dissipative optical traps for neutral atoms as an alternative to MOT.

\section{Kinetics of lithium atoms in a resonant bichromatic field}
\begin{figure}[h]
\centerline{\includegraphics[width=2.7 in]{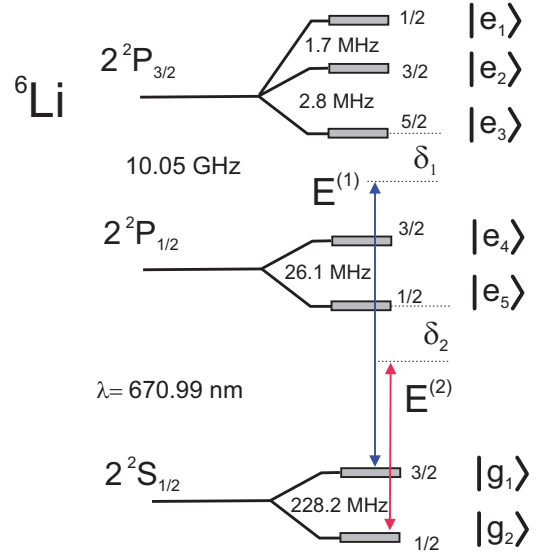}} \caption{Level
structure of $^{6}$Li atom.} \label{fig:Li6}
\end{figure}

In this section, we describe  the kinetics of lithium atoms in a bichromatic light field with frequencies $\omega_1$ and $\omega_2$
\begin{equation}\label{field0}
{\bf E}({\bf r},t) =\mbox{Re}\left\{ {\bf E}^{(1)}({\bf r}) e^{-i \omega_1 t}+{\bf E}^{(2)}({\bf r}) e^{-i \omega_2 t} \right\} \, ,
\end{equation}
where vector amplitudes ${\bf E}^{(1)}$ and ${\bf E}^{(2)}$  are determined by a superposition of running waves
\begin{equation}\label{rwavefield}
  {\bf E}^{(n)}({\bf r})  = \sum_m {\bf E}^{(n)}_m e^{i {\bf k}^{(n)}_m \cdot {\bf r}},\,\,\, n=1,2\, .
\end{equation}
The directions of the waves propagation are given by  wave vectors ${\bf k}^{(n)}_m$.
The frequencies of the light field components $\omega_1$ and $\omega_2$ are near the resonance lines $D_2$ and $D_1$ of lithium atoms, respectively. The scheme of energy levels and optical transitions used for laser cooling of $^6$Li atoms is shown in Fig.\,\ref{fig:Li6}.

It should be noted that the hyperfine splitting of the $^2P_{3/2}$ and $^2P_{1/2}$ states for lithium atoms is comparable to the natural linewidth  $\gamma/2\pi = 5.9$ MHz.  Moreover, the hyperfine components of the $^2P_{3/2}$ state are within the natural linewidth $\gamma$, which requires the use of a model taking into account the interaction of light waves with all hyperfine components.  Further, to simplify notation, we introduce the indices $e_\alpha$ ($\alpha = 1..5$) for hyperfine components of the excited states $^2P_{3/2}$ and $^2P_{1/2}$, and $g_n$ ($n = 1,2$) for the hyperfine components of the ground state $^2S_{1/2}$ (see Fig. \ref{fig:Li6}). Note that $^7$Li and $^6$Li isotopes have a similar level structure. Here we perform our analysis for the  $^6$Li atoms, that have lower values of total angular momentum. However, the results can also be generalized for $^7$Li atoms as well.

The kinetics of atoms in light fields is described by  the quantum kinetic equation for the atomic density matrix
\begin{equation}\label{master_eq}
    \frac{\partial }{\partial t} {\hat \rho}=-\frac{i}{\hbar}\left[{\hat H}_{kin} +{\hat H}_{int} + {\widehat W}(t), {\hat \rho}\right] +{\hat \Gamma} \left\{ {\hat \rho} \right\} \, ,
\end{equation}
where  ${\hat H}_{ kin} = {\hat p}^2/2M$ is the kinetic part, ${\widehat W}(t) = -{\hat {\bf d}} \cdot {\bf E(t)}$ describes interaction with the field in the dipole approximation (where ${\hat {\bf d}}$ is the dipole moment operator), and ${\hat H}_{int} $ is the Hamiltonian of a free atom in the rest frame
\begin{equation}
{\hat H}_{int} = \sum_{\alpha = 1 \dots 5} {\cal E}_{e_{\alpha}} {\hat P}^{(e_{\alpha})} + \sum_{n = 1,2} {\cal E}_{g_{n}} {\hat P}^{(g_{n})} \, ,
\end{equation}
where ${\cal E}_{e_{\alpha}}$ and ${\cal E}_{g_{n}}$ are energies of excited states $| e_{\alpha} \rangle$ and ground states $| g_{n} \rangle$. Here ${\hat P}^{(e_{\alpha})}$ and ${\hat P}^{(g_{n})}$ are
projection operators onto the Zeeman components of the hyperfine level $|e_{\alpha} \rangle $ and $|g_n \rangle $
\begin{eqnarray}
  {\hat P}^{(e_\alpha)} &=& \sum_{\mu_{e_\alpha} =-F_{e_\alpha} }^{ F_{e_\alpha}}|F_{e_\alpha}, \mu_{e_\alpha} \rangle \langle F_{e_\alpha}, \mu_{e_\alpha}|,\,\,\; \alpha = 1\dots 5   \nonumber \\
  {\hat P}^{(g_n)} &=& \sum_{\mu_{g_n} =-F_{g_n} }^{ F_{g_n}}|F_{g_n}, \mu_{g_n} \rangle \langle F_{g_n}, \mu_{g_n}|,  \,\,\; n = 1,2 .
\end{eqnarray}
The operator ${\hat \Gamma} \left\{ {\hat \rho} \right\}$ describes the relaxation of an atom due to spontaneous emission.

To use the resonant approximation in a bichromatic field (\ref{field0}), let's introduce the operator
\begin{equation}\label{rwa}
    {\hat T} = \mbox{exp}\left[-i\,t \left(\sum_{\alpha=1,2,3} {\omega_1}\, {\hat P}^{(e_\alpha)} + \sum_{\alpha=4,5} {\omega_2}\, {\hat P}^{(e_{\alpha})}\right) \right],
\end{equation}
where the summation is taken over all components of the levels $^2P_{3/2}$ and $^2P_{1/2}$. For the transformation (\ref{rwa}), the evolution equation for the transformed density matrix ${\hat {\tilde \rho}} = {\hat T}^{\dag}{\hat \rho} \,{\hat T}$ takes a similar form to (\ref{master_eq})
\begin{equation}\label{master_eqm}
    \frac{\partial }{\partial t} {\hat {\tilde \rho}}=-\frac{i}{\hbar}\left[{\hat H}_{kin} +{\widehat {\widetilde H}}_{int} + {\widehat {\widetilde W}}, {\hat {\tilde \rho}}\right] +{\hat \Gamma} \left\{ {\hat {\tilde \rho}} \right\} \, .
\end{equation}
However, the resonance approximation allows to eliminate the time dependence in the interaction operator and divide it into the sum of contributions
\begin{equation}\label{W}
   {\widehat {\widetilde W}} = \hbar {\hat V}^{(1)} + \hbar {\hat V}^{(2)} + h.c. \, ,
\end{equation}
determined by the interaction with two frequency components: ${\bf E}^{(1)}$ is close to the resonance with $D_2$ line, and ${\bf E}^{(2)}$ is close to the resonance with $D_1$  line (Fig.\ref{fig:Li6})
\begin{eqnarray}
{\hat V}^{(1)} &=& {\hat V}^{e_1g_1}+{\hat V}^{e_2g_1}+{\hat V}^{e_3g_1} \nonumber \\
{\hat V}^{(2)} &=& {\hat V}^{e_4g_2}+{\hat V}^{e_5g_2} \, .
\end{eqnarray}
Here, the operator blocks ${\hat V}^{e_{\alpha} g_n}$ are defined as
\begin{equation}
    {\hat V}^{e_{\alpha} g_n} = - \left( {\bf E}^{(n)} \cdot {\bf \hat D}^{e_{\alpha}g_n} \right) {\bar d}/\hbar \, ,
\end{equation}
where ${\bar d}$ is the reduced matrix element of the dipole moment operator, and ${\bf \hat D}^{e_{\alpha}g_n}$ are the matrix blocks of dipole moment operator
\begin{equation}\label{Dop}
{\bf {\hat d}} = {\hat {\bf D}} \,{\bar d}  + h.c. \, .
\end{equation}
Thus the the matrix blocks ${\bf \hat D}^{e_{\alpha}g_n}$ are
\begin{equation}
    {\bf \hat D}^{e_{\alpha}g_n} = {\hat P}^{(e_\alpha)}\, {\hat {\bf D}} \;{\hat P}^{(g_n)}\, .
\end{equation}
The decomposition of dipole moment operator ${\bf \hat D}$ in circular basis
\begin{equation}
{\hat {\bf D}} = \sum_{q=0,\pm1} {\hat D}_q \,{\bf e}^q \, ,
\end{equation}
are determined by the corresponding matrix elements $\langle I, J_{e_\alpha}; F_{e_\alpha},\mu_{e_\alpha} |{\hat D}_q|I, J_{g_n}; F_{g_n},\mu_{g_n} \rangle $ for atomic states with electronic angular momenta of the ground and excited states: $J_{g_n} = 1/2$ and $J_{e_\alpha} = 1/2,\, 3/2$ (Fig.\,\ref{fig:Li6}).
Here $I$ is the nuclear spin ($I= 1$ for $^6$Li atoms). According to the Wigner-Eckart theorem \cite{Varshalovich}, matrix elements are expressed through Clebsch-Gordan coefficients and 6-j symbols
\begin{widetext}
\begin{equation}
 \langle I, J_{e_\alpha}; F_{e_{\alpha}},\mu_{e_{\alpha}} |{\hat D}_q|I, J_{g_n}; F_{g_n},\mu_{g_n} \rangle  =
 C^{F_{e_{\alpha}}, \mu_{e}}_{F_{g_n},\mu_{g_n};\,1,q}
\, (-1)^{(J_{e_\alpha}+F_{g_n}+I+1)}\sqrt{(2F_{g_n}+1)(2J_{e_{\alpha}}+1)}
\left\{%
\begin{array}{ccc}
J_{e_\alpha} & 1 & F_{g_n} \\
J_{g_n} & I & J_{e_\alpha} \\
\end{array}%
\right\}.
\end{equation}
\end{widetext}

The non-Hamiltonian evolution of the system due to  spontaneous emission of the light field photons is described by ${\hat \Gamma} \left\{ {\hat {\tilde \rho}} \right\}$ in the equation for the density matrix (\ref{master_eq}) and (\ref{master_eqm}). Taking into account the recoil effects, this contribution has the form (see, for example, \cite{pru2007,pru2011})
\begin{equation}
 {\hat \Gamma} \left\{ {\hat {\tilde \rho}} \right\} = -\frac{\gamma}{2}\left({\hat P}^{(e)}{\hat {\tilde \rho}} +{\hat {\tilde \rho}}\, {\hat P}^{(e)}  \right) + {\hat \gamma}\left\{ {\hat {\tilde \rho}} \right\} \, ,
\end{equation}
with
\begin{eqnarray}\label{gammasmall}
&&{\hat \gamma}\left\{ {\hat {\tilde \rho}} \right\} = \gamma \frac{3}{2} \times  \\
&&\left< \sum_{\xi=1,2}{\left( {\hat {\bf D}} \cdot {\bf e}_{\xi}({\bf k}_{\xi}) \right)^{\dagger}} e^{-i {\bf k}_{\xi} \cdot {\bf {\hat r}}} {\hat {\tilde \rho}}\, e^{i {\bf k}_{\xi} \cdot {\bf {\hat r}}}  \left( {\hat {\bf D}} \cdot {\bf e}_{\xi}({\bf k}_{\xi}) \right) \right>_{\Omega_{\xi}}  . \nonumber
\end{eqnarray}
Here the operator ${\hat P}^{(e)} = \sum_\alpha{{\hat P}^{(e_\alpha)}}$ is a projector onto excited states (the summation is taken over all states of the $^2P_{3/2}$ and $^2P_{1/2}$ levels), $\left< \dots \right>_{\Omega_{\xi}}$ means averaging over the angles of spontaneously emitted photons with two orthogonal polarizations ${\bf e}_{\xi}$, ($\xi = 1,2$), the wave vectors ${\bf k}_{\xi}$ are specifying the direction of spontaneously emitted photon, and ${\bf {\hat r} }$ is position operator.

The Hamiltonian of a free atom in the rest frame in the rotating basis (\ref{rwa}) takes the form:
\begin{equation}\label{Hint}
   {\widehat {\widetilde H}}_{int} = - \hbar \sum_{\alpha=1..5} {\Delta_\alpha}\, {\hat P}^{(e_\alpha)} \, ,
\end{equation}
where
\begin{eqnarray}
    \Delta_{\alpha} &=& \omega_1 - \omega_{e_{\alpha}g_1},\,\mbox{for}\,\, \alpha=1,2,3 \nonumber \\
    \Delta_{\alpha} &=& \omega_2 - \omega_{e_{\alpha}g_2},\,\mbox{for}\,\, \alpha=4,5 \,,
\end{eqnarray}
are detunings from resonance for light induced transitions between hyperfine levels of $D_2$ and $D_1$ lines caused by two frequency components  Fig.\,\ref{fig:Li6}. Here $\omega_{e_{\alpha}g_n} = ({\cal E}_{e_\alpha}-{\cal E}_{g_n})/\hbar $ is the transition frequency between states $|e_\alpha \rangle$ and $|g_n \rangle$.

Note, that the kinetics of lithium atoms is characterized by enough small recoil parameter $\varepsilon_R = E_R/\hbar \gamma \simeq 0.01$ ($E_R = \hbar^2 k^2/2M$ is recoil energy), that makes it possible to use the semiclassical approximation. Indeed, for such a small recoil parameter, the semiclassical approximation gives a good agreement with quantum treatment based on direct numerical solution  of the density matrix equation (\ref{master_eqm}) in monochromatic light \cite{kirp2022,kirp2020}. Within the semiclassical approximation, the equation for the density matrix (\ref{master_eqm}) can be reduced to the Fokker-Planck equation (see, for example \cite{dal85sem,Javanainen91,pru99})
\begin{eqnarray}\label{FPequation}
    &&\left(
    \frac{\partial}{\partial t} + \frac{{\bf p}}{M} \cdot \nabla \right){\cal F} = \nonumber \\
    &&- \sum_i \frac{\partial}{\partial p_i} f_i({\bf r}, {\bf p}) {\cal F} +\sum_{i,j} \frac{\partial^2}{\partial p_i\, \partial p_j} D_{ij}({\bf r}, {\bf p}) {\cal F} \,,
\end{eqnarray}
for the distribution function of atoms in the phase space
${\cal F}({\bf r}, {\bf p}) = Tr\{ {\hat \rho} ({\bf r}, {\bf p})\}$, where the trace is taken over internal degrees of atom density matrix in the Wigner representation. Here $\nabla$ is  spatial gradient, $f_i$ are the Cartesian components of the light force on atoms, and $D_{ij}$ are the Cartesian components of the diffusion tensor in momentum space. The force and diffusion tensor can be obtained in a process of reduction of the density matrix equation (\ref{master_eq}) to the Fokker-Plank equation (\ref{FPequation}).

\section{Bichromatic optical lattice for lithium atoms}
The light force on atom in a light field is determined by the spatial gradients of the interaction operator (\ref{W})
\begin{equation}\label{Force}
    f_i({\bf r}, {\bf v}) = -Tr\left\{ \bigg(\nabla_i \, {\widehat {\widetilde W}}({\bf r}) \bigg)\, {\hat \sigma}({\bf r}, {\bf v}) \right\} \, ,
\end{equation}
where ${\hat \sigma}({\bf r}, {\bf v})$ is the stationary solution of the equation (\ref{master_eqm}) in zero order by recoil effects
\begin{eqnarray}\label{master_eq0}
    &&  ({\bf v} \cdot \nabla) \,{\hat \sigma}({\bf r}, {\bf v})= \nonumber \\
    &&-\frac{i}{\hbar}\left[{\widehat {\widetilde H}}_{int}+{\widehat {\widetilde W}}({\bf r}), {\hat \sigma}({\bf r}, {\bf v}) \right] +{\hat \Gamma}^{(0)} \left\{ {\hat \sigma}({\bf r}, {\bf v}) \right\} \, ,
\end{eqnarray}
with the normalization condition $\mbox{Tr}\left\{ {\hat \sigma} \right\} = 1$. The operator of spontaneous relaxation in zeroth order by recoil is
\begin{eqnarray}
    {\hat \Gamma}^{(0)} \left\{ {\hat \sigma} \right\} &=& -\frac{\gamma}{2}\left({\hat P}^{(e)}{\hat \sigma}+{\hat \sigma}\, {\hat P}^{(e)}  \right) + {\hat \gamma}^{(0)}\left\{ {\hat \sigma} \right\} \nonumber \\
{\hat \gamma}^{(0)}\left\{ {\hat \sigma} \right\} &=& \gamma \sum_{q=0,\pm 1} {\hat  D}_q^{\dagger} \,{\hat \sigma} \,{\hat  D}_q \, .
\end{eqnarray}
Let us note, the expression (\ref{Force}) define general expression for the force is a function of atom position ${\bf r}$ and velocity ${\bf v}$. The optical lattice, the optical potential for atoms is determined by the force on moveless atom $v=0$. The expression for the force can be obtained numerically based on the solution of (\ref{master_eq0}) for the density matrix $ {\hat \sigma}$, however, analytical solutions are of the greatest interest for analysis.

The force, by its nature, is divided into two components: the induced force, associated with the transmission of momentum from light field to an atom in the processes of stimulated absorption/emission of photons between different spatial modes of the field, and the spontaneous light pressure force, associated with the transfer of momentum to atoms in the cycle of induced absorption and spontaneous emission of photons. Both components of the force can lead to the formation of an optical potential, or the so-called optical lattice \cite{metcalf,jessen96,Grynberg,bez2005}.

For analysis of optical potential in bichromatic field, we can get analytical
expression for the force on slow atoms $kv < \gamma$ in  the light fields of low intensity
\begin{equation}\label{slow}
    S^{(n)}= \frac{|\Omega^{(n)}({\bf r})|^2}{\gamma^2 + 4 \delta_n^2} \ll 1, \,\,\, n=1,2 \, ,
\end{equation}
where
\begin{equation}\label{Rabi}
    \Omega^{(n)}({\bf r}) = -\frac{|{\bf E}^{(n)}({\bf r})| {\bar d}}{\hbar} \, ,
\end{equation}
is the local Rabi frequency for $n$-th frequency component (\ref{rwavefield}),
\begin{eqnarray}
    \delta_1 &=& \omega_1 - \omega_{e_3g_1} \nonumber \\
    \delta_2 &=&  \omega_2 - \omega_{e_5 g_2} \, ,
\end{eqnarray}
are main detunings from the resonances, which are determined by the frequency $\omega_{e_3g_1}$ of the transition $^2P_{3/2}(F=5/2) \to\, ^2S_{1/2}(F=3/2)$ for component ${\bf E}^{(1)}$, and the  frequency $\omega_{e_5g_2}$ of the transition  $^2P_{1/2}(F=1/2) \to \,^2S_{1/2}(F=1/2)$ for component ${\bf E}^{(2)}$ as shown in Fig.\,\ref{fig:Li6}.

In the limit (\ref{slow}), the populations of the excited states are small, and  for slow atoms $kv < \gamma$ the equation (\ref{master_eq0}) can be reduced to the equation for the density matrices of two ground states $^2S_{1/2}$ with angular momenta $F=3/ 2$ and $F=1/2$
\begin{equation}
{\hat \sigma}^{(n)} = {\hat P}^{(g_n)} {\hat \sigma}\, {\hat P}^{(g_n)},\,\, n=1,2 \, .
\end{equation}
In this case, the  equations for these  matrices take the form
\begin{eqnarray}\label{reduced}
( {\bf v} \cdot \nabla)\,&{\hat \sigma}^{(1)} & \,= \nonumber \\
  &-&i \left[{\hat H}_{eff}^{(1)},{\hat \sigma}^{(1)} \right] + {\hat P}^{(g_1)}\, {\hat \gamma}^{(0)}\left\{  {\hat \sigma}^{ee} \right\} \, {\hat P}^{(g_1)}\,, \nonumber \\
( {\bf v} \cdot \nabla)\,&{\hat \sigma}^{(2)}& \,= \\
&-&i \left[{\hat H}_{eff}^{(2)}, {\hat \sigma}^{(2)} \right] + {\hat P}^{(g_2)}\, {\hat \gamma}^{(0)}\left\{  {\hat \sigma}^{ee} \right\} \, {\hat P}^{(g_2)} \nonumber \, .
\end{eqnarray}
The Hamiltonian evolution here  is described by two effective Hamiltonians
\begin{eqnarray}\label{Hefff}
{\hat H}_{eff}^{(1)} & =& \sum_{\alpha=1,2,3}
\frac{\Delta_{\alpha}-i \gamma/2}{|\nu_{\alpha}|^2} \, \left({\hat V}^{e_{\alpha} g_1}\right)^{\dagger} {\hat V}^{e_{\alpha} g_1} \,,\nonumber \\
{\hat H}_{eff}^{(2)} & =&\sum_{\alpha =4,5} \frac{\Delta_{\alpha}-i \gamma/2}{|\nu_{\alpha}|^2} \,\left({\hat V}^{e_{\alpha} g_2}\right)^{\dagger} {\hat V}^{e_{\alpha}g_2} \,.
\end{eqnarray}
Here we use the notation $\nu_{\alpha} = \gamma/2 - i\,\Delta_{\alpha}$. The density matrix of excited states ${\hat \sigma}^{ee}$ in the zeroth order by recoil for slow atoms is divided into a sum of two blocks
\begin{eqnarray}\label{reducedeee}
{\hat \sigma}^{ee} &=& {\hat \sigma}^{ee}_1 + {\hat \sigma}^{ee}_2  \nonumber \\
{\hat \sigma}^{ee}_1 &=& \sum_{{\alpha},{\alpha'} = 1,2,3} \frac{1}{ \nu_{\alpha} \, \nu_{\alpha'}^* }\, {\hat V}^{e_{\alpha}g_1}\, {\hat \sigma}^{(1)} \, \left({\hat V}^{e_{\alpha'}g_1} \right)^{\dagger}    \nonumber \\
{\hat \sigma}^{ee}_2 &=& \sum_{{\alpha},{\alpha'} = 4,5} \frac{1}{ \nu_{\alpha} \, \nu_{\alpha'}^* }\, {\hat V}^{e_{\alpha}g_2} \,{\hat \sigma}^{(2)}  \left({\hat V}^{e_{\alpha'}g_2} \right)^{\dagger}  \, .
\end{eqnarray}
Note, that for sufficiently large detunings that exceed the hyperfine splitting in excited states $^2P_{3/2}$ and $^2P_{1/2}$ and the natural width $\gamma$, i.e.
\begin{eqnarray}\label{bigdelta}
 &\delta_1&\,\gg ( \omega_{e_1e_2},\, \omega_{e_2e_3},\, \gamma ) \nonumber \\
 &\delta_2& \,\gg (\omega_{e_4e_5},\, \gamma) \, ,
\end{eqnarray}
the following approximation can be used
\begin{eqnarray}\label{delta12}
 \Delta_{\alpha}& \simeq& \delta_1 \,\,\, \mbox{for} \,\,{\alpha}=1,2,3 \nonumber \\
\Delta_{\alpha} & \simeq & \delta_2  \,\,\, \mbox{for} \,\,{\alpha}=4,5 \, .
\end{eqnarray}
In this case, the effective Hamiltonians (\ref{Hefff}) are substantially simplified and reduced to the shift operators for the hyperfine components of the ground states
\begin{eqnarray}\label{Un}
{\hat H}_{eff}^{(n)}({\bf r}) &=& \delta_n\, S^{(n)}({\bf r}) \,\, {\hat U}^{(n)}({\bf r}), \,\,\, n=1,2  \\
{\hat U}^{(1)}({\bf r}) &=& \frac{2}{3} |{\bf A}_1({\bf r})|^2 \,{\hat P}^{(g_1)} -  \frac{i}{3} \bigg( \left[ {\bf A}_1 ^{*}({\bf r}) \times {\bf A}_1({\bf r})  \right] \cdot \frac{\hat {\bf F}}{F_{g_1}} \bigg), \nonumber \\
{\hat U}^{(2)}({\bf r}) &=& \frac{1}{3} |{\bf A}_2({\bf r})|^2 \,{\hat P}^{(g_2)} -  \frac{i}{9} \bigg(\left[ {\bf A}_2 ^{*}({\bf r}) \times {\bf A}_2({\bf r}) \right] \cdot \frac{{\hat {\bf F}}}{F_{g_2}} \bigg), \nonumber
\end{eqnarray}
are determined by the spatial configuration of local polarization vectors for each frequency component: ${\bf A}_1({\bf r}) = {\bf E}^{(1)}({\bf r})/|{\bf E}^{(1)}({\bf r})|$ and ${\bf A}_2({\bf r}) = {\bf E}^{(2)}({\bf r})/|{\bf E}^{(2)}({\bf r})|$. Here ${\bf \hat F}$ is the total angular momentum operator, and  $\left[{\bf A} \times {\bf B} \right]$ denotes the cross product of the vectors ${\bf A}$ and ${\bf B}$.  The expressions (\ref{Un}) are written in the invariant form and valid for an arbitrary nuclear spin (i.e., they are also applicable to $^7$Li atoms). Note that the expression for ${\hat U}^{(1)}$ corresponds to optical lattice light shift in \cite{jessen98,jessen98_2} obtained for monochromatic light field far detuned to optical resonance $D_2$ line of Cs atoms.

In the considered limit (\ref{slow}), the force on  atoms can be represented as sum of two parts
\begin{equation}\label{forcelow}
    {\bf f} = \hbar \,\mbox{Tr}\left\{ {\bf {\hat f}}_1\, {\hat \sigma}^{(1)} \right\} + \hbar \,\mbox{Tr}\left\{ {\bf {\hat f}}_2 \,{\hat \sigma}^{(2)} \right\}\,,
\end{equation}
where the force vector operators ${\bf {\hat f}}_n$ are determined by the spatial gradients
\begin{equation}
 {\bf {\hat f}}_n = - \nabla {\hat H}_{eff}^{(n)}\,,\,\, n=1,2\,.
\end{equation}
It should be noted that a distinctive feature of the considered bichromatic configuration, compared to optical lattices formed by monochromatic fields, is the presence of an additional shift ${\hat U}^{(2)}$ with a spatial dependence that may be different from ${\hat U}^{(1)}$. Spatially non-uniform optical pumping of the ground state levels $|g_1 \rangle$ and $|g_2 \rangle$, defined by equation (\ref{reduced}) under the motion in two potentials, leads to the rectification of the dipole force (\ref{forcelow}) on the wavelength scale. In result, the optical potential with a macroscopic period can be formed. To demonstrate this,  we consider several examples bellow.

\subsection{One-dimensional optical lattices created by a bichromatic field}
Let us consider the general form of a one-dimensional bichromatic field configuration created by counter-propagating waves of equal scalar amplitude along the $z$ axis
\begin{equation}\label{lin0}
{\bf E}^{(n)}(z) =  E^{(n)}_0\left({\bf A}^{(n)}_+ e^{i k^{(n)} z} + {\bf A}^{(n)}_- e^{-i k^{(n)} z}\right),\,\,\, n =1,2 \, .
\end{equation}
Here $E^{(n)}_0$ are scalar amplitudes, and ${\bf A}^{(n)}_{\pm}$ are unit polarization vectors of the counter-propagating light waves of different frequency components ($n=1,2$), which can be elliptical in general case. If  we denote $k^{(2)} = k$, and $\Delta k = k^{(1)}-k^{(2)}>0$, then the field (\ref{lin0}) can be rewritten as:
\begin{eqnarray}
  {\bf E}^{(1)}(z) &=&  E^{(1)}_0\left({\bf A}^{(1)}_+ e^{i kz + i\Delta \phi} + {\bf A}^{(1)}_- e^{-i kz- i\Delta \phi}\right)  \, ,\nonumber \\
   {\bf E}^{(2)}(z) &=&  E^{(2)}_0\left({\bf A}^{(2)}_+ e^{i k z} + {\bf A}^{(2)}_- e^{-i k z}\right).
\end{eqnarray}
Since $\Delta k = k^{(1)}-k^{(2)} \ll  k$, the relative spatial phase
\begin{equation}
\Delta \phi = \Delta k\, z  \, ,
\end{equation}
can be considered constant over the wavelength scale $\lambda = 2\pi/k$.
The spatial polarization configurations of different frequency components  are determined by the mutual spatial orientation of the polarization vectors ${\bf A}^{(n)}_{+}$ and ${\bf A}^{(n)}_{-}$.  Bellow we consider several configurations.

\subsubsection{Double $lin\,||\,lin$ field configuration}
\begin{figure}[b]
\centerline{\includegraphics[width=2.6 in]{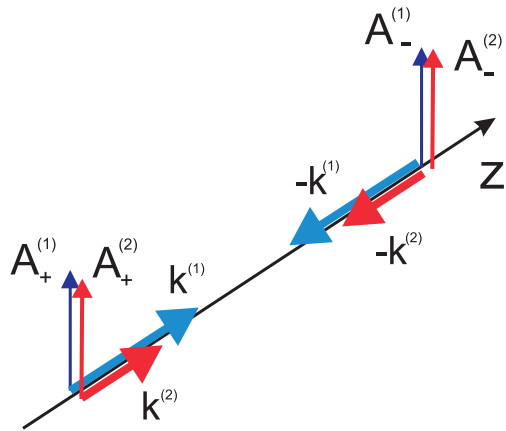}}
\caption{Double $lin\,||\,lin$ field configuration}
\label{fig:linfield}
\end{figure}

For a bichromatic field created by standing waves with identical linear polarizations ${\bf A}^{(1)}_{\pm} = {\bf A}^{(2)}_{\pm}={\bf A} = {\bf A}^*$ (the so-called double $lin\,||\,lin$ configuration, see Fig.\ref{fig:linfield}), the effective Hamiltonians (\ref{Un}) in the limit of large detunings (\ref{bigdelta}) are reduced to the scalar light shifts $u_1 = 2\delta_1 S_1 |{\bf A}|^2/3$ and $u_2 = \delta_2 S_2 |{\bf A}|^2/3$ with a spatial dependence represented by standing waves of each frequency components
\begin{eqnarray}\label{Hefflin0}
{\hat H}_{eff}^{(n)} & =& u_n\, {\hat P}^{(g_n)},\,\mbox{for} \,\,\; n=1,2 \nonumber \\
u_1 &=&  \frac{8}{3}\, \delta_1\, s_1 \,\cos^2\left(kz + \Delta\phi \right)\nonumber \\
u_2 &=&  \frac{4}{3} \,\delta_2\, s_2 \,\cos^2\left(kz\right)\,.
\end{eqnarray}
Here $s_1$ and $s_2$ are the saturation parameters per one running wave, $s_n = |\Omega_0^{(n)}|^2/(4\delta_n^2+\gamma^2)$, where $\Omega_0^{(n)} = -E_0^{(n)} {\bar d}/\hbar$ are the corresponding Rabi frequencies. The optical potential of the light field is determined by the force (\ref{Force}), (\ref{forcelow}) on atom at rest. The solution of Eq.\,(\ref{reduced}) for $v = 0$ in this field leads to an isotropic distribution on each of the ground levels  $|g_1 \rangle$ and $|g_2 \rangle$
\begin{eqnarray}\label{sigmalin0}
    {\hat \sigma}^{(1)} &=& \frac{\cos^2(kz)\,s_2}{2s_1\cos^2(kz+\Delta \phi)+4s_2\cos^2(kz)} \, {\hat P}^{(g_1)} \,, \nonumber \\
    {\hat \sigma}^{(2)} &=& \frac{\cos^2(kz+\Delta \phi)\,s_1}{2s_1\cos^2(kz+\Delta \phi)+4s_2\cos^2(kz)} \, {\hat P}^{(g_2)} \, ,
\end{eqnarray}
since
$\left[ {\bf A}_n^* \times {\bf A}_n \right]=0$ in (\ref{Un}) for this field configuration Fig.\ref{fig:linfield}.
Thus, the force (\ref{forcelow}) resulting to formation of an optical potential is divided into two contributions from each of the frequency components
\begin{eqnarray}\label{forcelin0}
f &= & f_1+f_2 \,,\nonumber \\
f_1 &=& \hbar k \frac{16}{3} \frac{ \delta_1 s_1 s_2\,\sin(2kz +2\Delta \phi)\cos^2(kz)}{s_1\cos^2(kz+\Delta \phi)+2s_2\cos^2(kz)} \,,\nonumber \\
f_2 &=& \hbar k \frac{4}{3}  \frac{\delta_2 s_1 s_2\,\sin(2kz)\cos^2(kz+\Delta\phi)}{s_1\cos^2(kz+\Delta \phi)+2s_2\cos^2(kz)}\, .
\end{eqnarray}
\begin{figure}[t]
\centerline{\includegraphics[width=3.3 in]{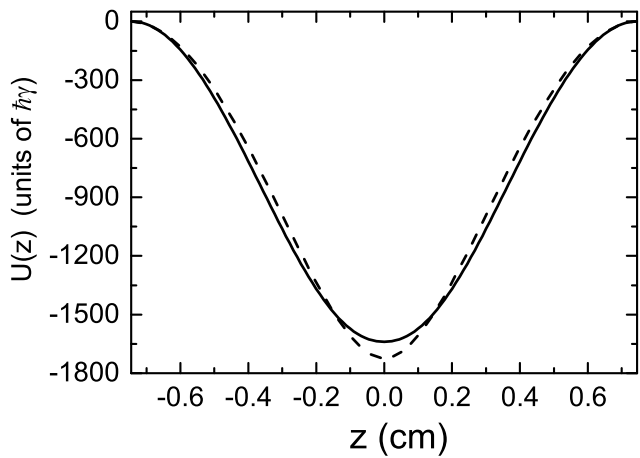}}
\caption{Macroscopic optical potential $U(z)$ in units of $\hbar
\gamma$ of a bichromatic lattice formed by field of double
$lin\,||\,lin$ configuration. The solid line represents the solution
based on the analytical expressions obtained in the approximation of
small saturation and large detunings (\ref{forcelin0}), while the
dotted line represents the optical lattice potential obtained by
direct numerical solution for the force (\ref{Force}), i.e. outside
the indicated approximations.   The detunings are $\delta_1 = -2
\gamma$ and $\delta_2 = -3 \gamma$, the saturation parameters are
$s_1 = s_2 = 0.1$ (i.e. Rabi per one wave  $\Omega_0^{(1)}  \simeq
1.3 \gamma$, $\Omega_0^{(2)} \simeq 1.9 \gamma$). The depth of
macroscopic optical potential for considered parameters reaches
$\Delta U = 1720\, \hbar \gamma$ that corresponds $\simeq 0.5$ K in
temperature units.} \label{fig:uzlin0}
\end{figure}
As was mentioned above, due to the different spatial dependencies of the light shifts $u_1$, $u_2$ and the spatially inhomogeneous optical pumping of $|g_1\rangle$ and $|g_2 \rangle$ states, the effect of force rectification on the wavelength scales appears. This rectification effect is similar in nature to those described for a two-level atom in a bichromatic field \cite{kaz87,ovch90,pru2013,pru2017}, as well as in monochromatic fields formed by counter-propagating waves with elliptical polarizations \cite{pru2001}.
Thus, the averaged over the wavelength force
\begin{eqnarray}\label{rectforce}
{\bar f}(\Delta\phi) = \frac{1}{\lambda} \int_{0}^{\lambda} f(z)\, dz = {\bar f}_1(\Delta\phi) +{\bar f}_2(\Delta\phi)\, ,
\end{eqnarray}
in the general case, is not equal to zero, and its magnitude and sign are determined by the relative spatial phase $\Delta\phi$
\begin{widetext}
\begin{eqnarray}\label{favlin0}
{\bar f}_1 &=& \hbar k \frac{16}{3}\frac{\delta_1 s_1\, s_2\,\, \sin(2\Delta\phi)}{\left(8 s_1s_2 \cos^2(\Delta\phi)+(2s_2-s_1)^2 \right)^2} \left[|\sin(\Delta\phi)|\sqrt{2s_1s_2} \left((2s_2+3s_1)(2s_2-s_1)-8s_2s_1\cos^2(\Delta\phi)\right)\right. \nonumber \\
 &&+ \left. 16 s_2^2s_1\cos^2(\Delta\phi) -s_1(6s_2+s_1)(2s_2-s_1)\right]\,, \nonumber \\
 {\bar f}_2 &=& \hbar k \frac{4}{3}\frac{\delta_2 \,s_1\, s_2\,\, \sin(2\Delta\phi)}{\left(8 s_1s_2 \cos^2(\Delta\phi)+(2s_2-s_1)^2 \right)^2} \left[|\sin(\Delta\phi)|\sqrt{2s_1s_2} \left((6s_2+s_1)(2s_2-s_1)+8s_2s_1\cos^2(\Delta\phi)\right)\right. \nonumber \\
  &&- \left. 8 s_2s_1^2\,\cos^2(\Delta\phi) -2\,s_2\,(2s_2+3s_1)(2s_2-s_1)\right]\, .
\end{eqnarray}
\end{widetext}
As the result, on a distance exceeding the wavelength, the rectified force components ${\bar f}_1$ and ${\bar f}_2$ form a deep macroscopic potential
\begin{eqnarray}\label{u12lin0}
    U(z) &=& -\int {\bar f}_1\big(\Delta\phi(z) \big)\,dz-\int {\bar f}_2\big(\Delta\phi(z) \big)\, dz \nonumber \\ &=& U_1(z)+U_2(z)
\end{eqnarray}
with a period corresponding to change of the relative spatial phase $\Delta\phi$ from 0 to $\pi$, which for the lithium atoms in one-dimension field (\ref{lin0}) is $\Lambda \simeq 1.5$ cm (see Fig.\,\ref{fig:uzlin0}).
\begin{figure}[h]
\centerline{\includegraphics[width=3.0 in]{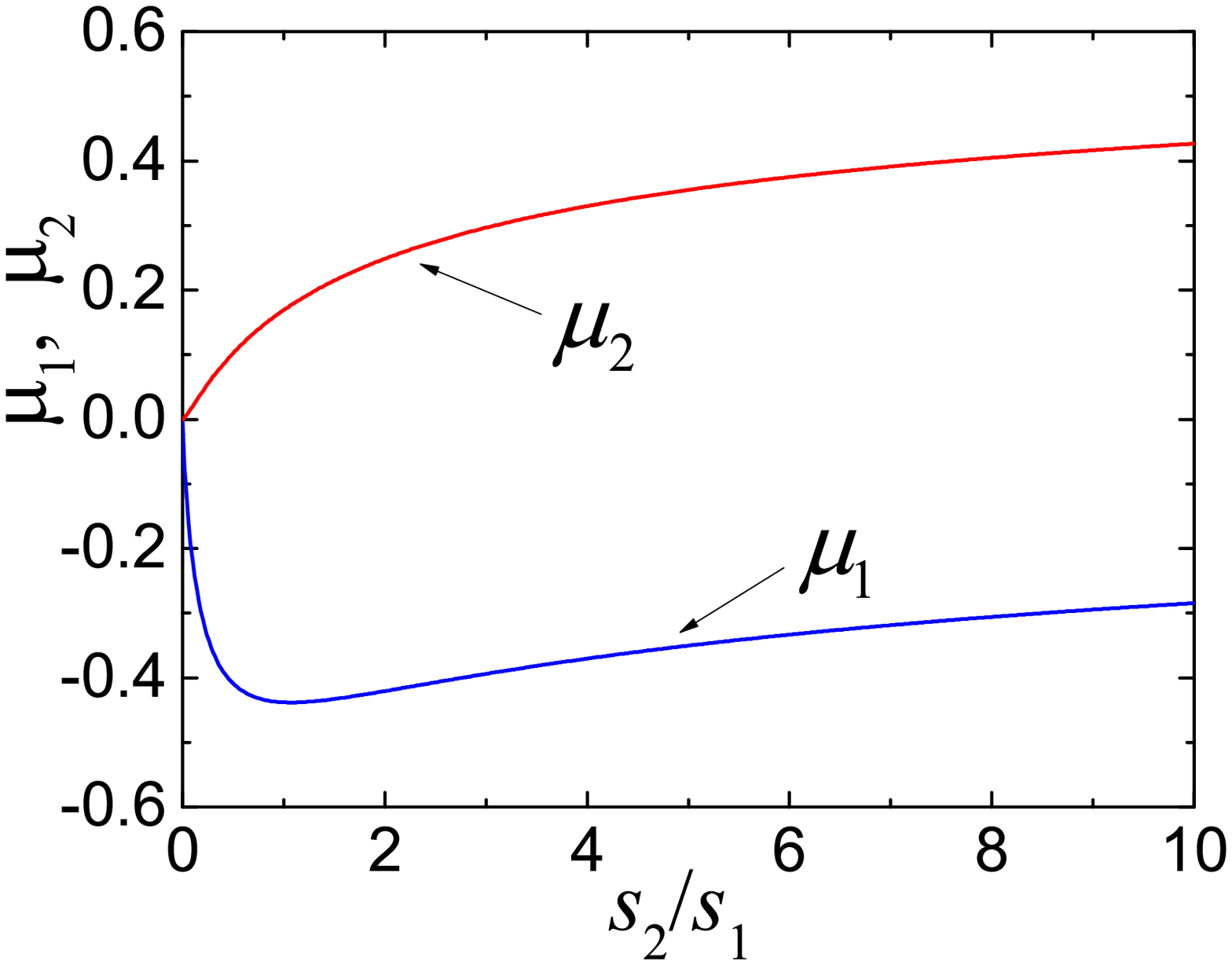}} \caption{The
functions $\mu_1(s_2/s_1)$ and $\mu_2(s_2/s_1)$, which determine the
depth of macroscopic potentials $U_1$ and $U_2$ in the field of
double $lin\,||\,lin$ configuration.} \label{fig:uu_LinLin}
\end{figure}
\begin{figure}[t]
\centerline{\includegraphics[width=3.0 in]{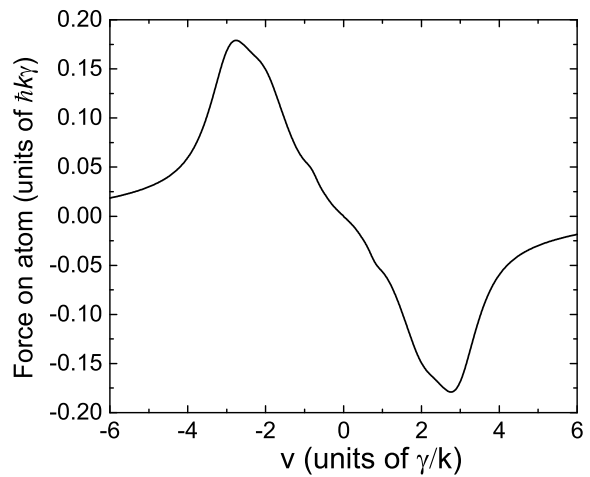}} \caption{The
force on lithium atoms in bichromatic field as function of velocity
in the region of the global minimum of the optical potential. The
field parameters correspond to Fig.\ref{fig:uzlin0}. }
\label{fig:Fv0}
\end{figure}
\begin{figure}[h]
\centerline{\includegraphics[width=3.4 in]{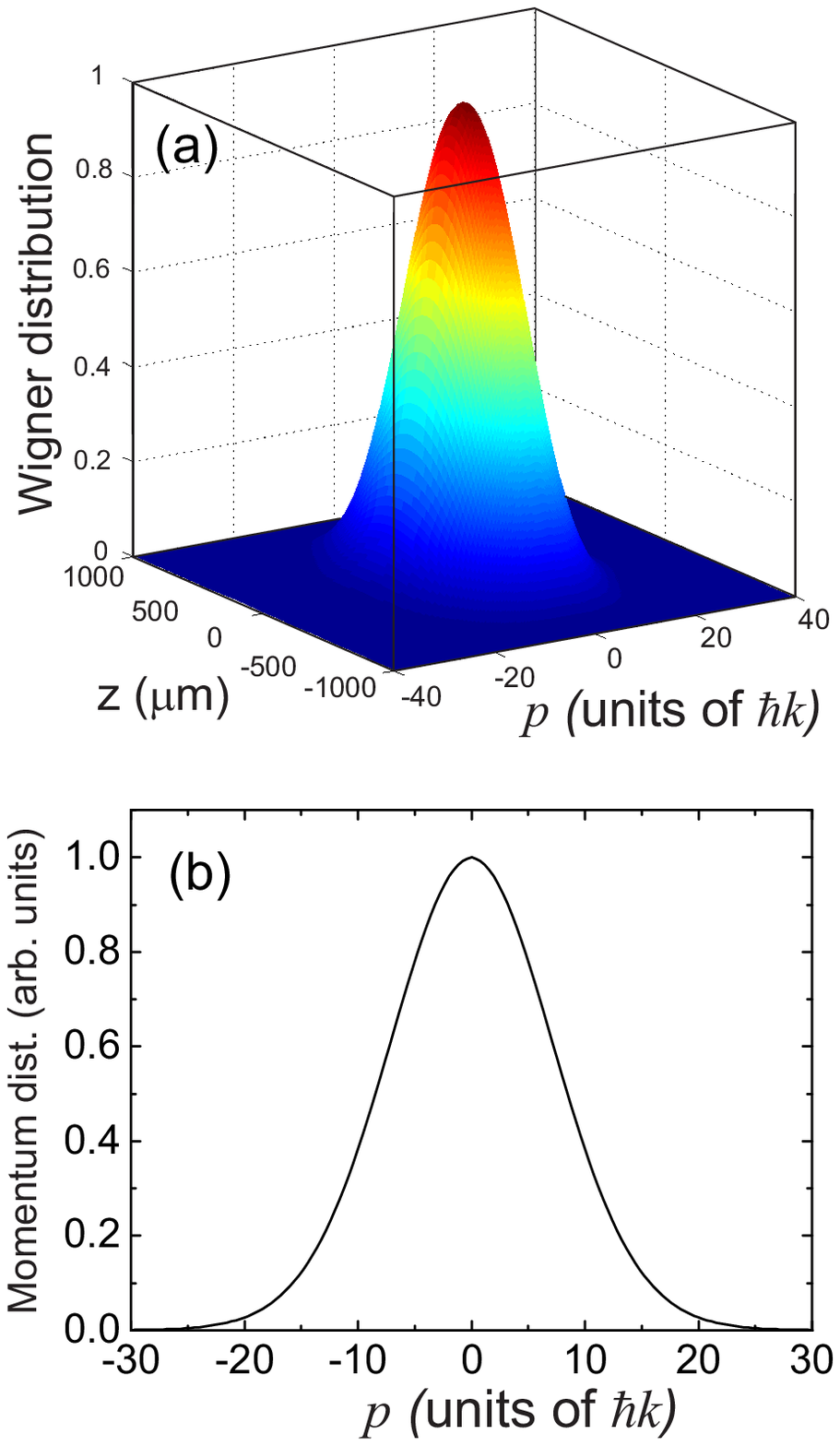}} \caption{The
Wigner function of the atomic phase distribution ${\cal F}(z,p)$ (a)
and the momentum distribution of cold atoms (b) in the macroscopic
potential of the bichromatic field of double $lin\,||\,lin$
configuration. The field parameters correspond to
Fig.\ref{fig:uzlin0}.} \label{fig:W1}
\end{figure}
In this case, the positions of the global maxima and minima of the potentials $U_n$ (\ref{u12lin0}) correspond to position where $\Delta\phi = 0, \pm \pi/2 , \dots$. The depth of each term in (\ref{u12lin0}) is determined by
\begin{equation}\label{DeltaU_n}
    \Delta U_n = \frac{1}{\Delta k}\int_{0}^{\pi/2} {\bar f}_n\left(\Delta\phi \right)\, d \Delta\phi , \,\, n=1,2 \, ,
\end{equation}
that results the following expressions for potentials depth
\begin{eqnarray} \label{u1u2_lin0}
\Delta U_1 &=& \hbar \delta_1 \frac{k}{\Delta k} s_1 \mu_1(s_2/s1)\,, \nonumber \\
\Delta U_2 &=& \hbar \delta_2 \frac{k}{\Delta k}  s_1 \mu_2(s_2/s1) \, .
\end{eqnarray}
The $\mu_1$ and $\mu_2$ here are dimensionless functions of the ratio $a = s_2/s_1$ (see Fig.\ref{fig:uu_LinLin})
\begin{eqnarray}
\mu_1(a) &=& -\frac{8}{3}\,a \left[
\ln{\left(
\frac{2\,a+1}{(\sqrt{2a}+1)^2}
\right)}+\frac{2}{\sqrt{2a}+1}
\right] \nonumber \\
\mu_2(a) &=& \frac{1}{3} \left[
\ln{\left(
\frac{2\,a+1}{(\sqrt{2a}+1)^2}
\right)}+\frac{\sqrt{8a}}{\sqrt{2a}+1}
\right] \, ,
\end{eqnarray}
which allow to estimate the depth of the macroscopic potential (\ref{u12lin0}) for various light field parameters.

The relatively large depth of the macroscopic potential, which significantly exceeds the magnitude of the light shifts of each frequency component, is provided by the multiplier $k/\Delta k$ (for lithium atoms $k/\Delta k\simeq 4.4\times 10^4$). As the result, for parameters of Fig.\,\ref{fig:uzlin0}, the depth of the macroscopic potential reaches $\simeq 0.5\, K$ in temperature units. However, for the localization of trapped atoms, the curvature at the points of minimum macroscopic potential plays a significant role. In the case of  bichromatic field it is proportional to $\Delta k/k$, that allows to localize the trapped Li atoms at sub-millimeter scales (see bellow).

For the bichromatic field with detunings $\delta_1$, $\delta_2$ near the resonant lines $D_2$ and $D_1$, the presence of dissipative Doppler force potentially allows trapping and cooling of atoms directly from the room temperature vapor. The Doppler mechanisms of laser cooling are result of imbalance of spontaneous light pressure forces from counter-propagating waves on  moving atoms \cite{metcalf} and acts on atoms over the all points of macroscopic potential. The Fig.\,\ref{fig:Fv0} shows dependence of the force on lithium atoms as function of atomic velocity. In particular, the force $f(v)$  for considered detunings $\delta_1 = -2 \gamma$ and $\delta_2 = -3 \gamma$ takes on maximum values for the velocity groups of atoms near resonance lines, near $|v| \simeq 3 \gamma/k$. As can be seen here the force has a sufficiently wide velocity range, comparable to the dissipative force in a standard MOT. The result for the force here is obtained on a base of eq. (\ref{Force}), i.e. beyond the limit of the low fields intensity (\ref{slow}) and slow atoms $kv < \gamma$.

The Fig.\,\ref{fig:W1} shows steady-state Wigner function ${\cal F}(z,p)$  and the momentum distribution of atoms in the trap, obtained by solving the Fokker-Planck equation (\ref{FPequation}) with taking into account the nonlinear dependence of the force (see Fig.\ref{fig:Fv0}) and diffusion coefficients on atoms velocity. For considered parameters, the momentum distribution of atoms is well approximated by a Gaussian distribution with temperature $T \simeq 1.1\, \hbar \gamma/k_B \simeq 300\, \mu K$ and corresponds to temperatures near the Doppler limit of laser cooling. The size of cold atom cloud in a macroscopic potential can be defined as
\begin{equation}\label{sizeZ}
    \beta_z = \sqrt{ \int z^2 {\cal F}(z,p)\, dz dp }\,\, ,
\end{equation}
where integration by coordinate is taken over the macroscopic period. For the parameters of Fig.\,\ref{fig:W1} the size of the atomic cloud  is  $\beta_z \simeq 320\, \mu m$.

\begin{figure}[t]
\centerline{\includegraphics[width=2.8 in]{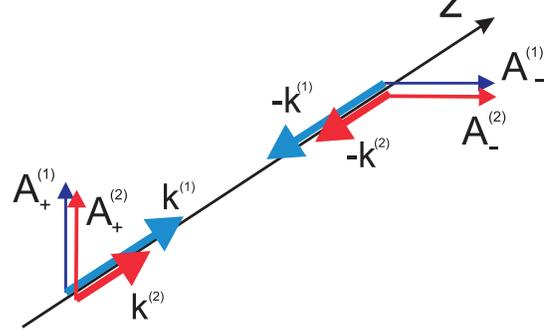}}
\caption{Double $lin$\,$\perp$\,$lin$ field configuration}
\label{fig:linfield1}
\end{figure}

\subsubsection{Double $lin$\,$\perp$\,$lin$ field configuration}

To achieve laser cooling below the Doppler limit, the light fields with polarization gradients are required. In our case, such a field can be formed with counter-propagating waves with different polarizations \cite{pru99,pru2001,Prudnikov1999}. For example, the well-known $lin$\,$\perp$\,$lin$ field  configuration, formed by waves with orthogonal linear polarizations, and $\sigma_{+} - \sigma_{-}$ configuration, formed by waves with opposite circular polarizations  \cite{dal1989}. However, the $\sigma_{+}- \sigma_{-}$ field does not lead to spatial modulation of light shifts according to (\ref{Un}) in the limit of large detunings (\ref{bigdelta}), and rectification effect is absent.

\begin{figure}[h]
\centerline{\includegraphics[width=3.0 in]{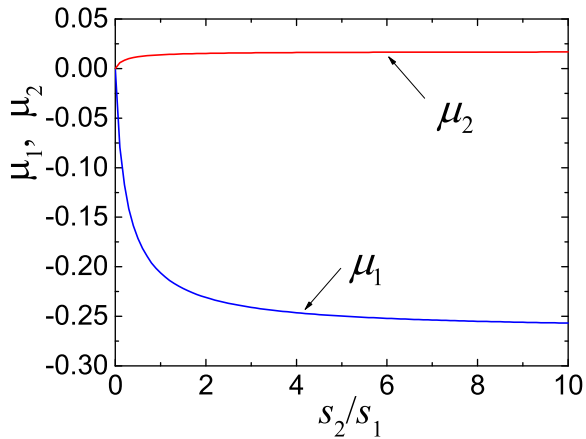}} \caption{The
functions $\mu_1(s_2/s_1)$ and $\mu_2(s_2/s_1)$, which determine the
depth of macroscopic potentials $U_1$ and $U_2$ in the field of
double $lin$\,$\perp$\,$lin$ configuration.} \label{fig:uu_LinPLin}
\end{figure}

In this section, we consider the double $lin$\,$\perp$\,$lin$  configuration (see Fig.\,\ref{fig:linfield1}), where ${\bf A}^{(1)}_{+} = {\bf A}^{(2)}_{+} = {\bf e}_x$  is unit vector along $x$-axis,  and ${\bf A}^{(1)}_{-} = {\bf A}^{(2)}_{-}={\bf e}_y$  is unit vector along $y$-axis
\begin{equation}
{\bf E}^{(n)}(z) = E^{(n)}_0 \left({\bf e}_x e^{ik^{(n)} z} + {\bf e}_y e^{-ik^{(n)} z}\right),\, n=1,2 \, .
\end{equation}
For this field, as well as for the double $lin || lin$ configuration, a deep macroscopic potential can be realized. The effective Hamiltonians (\ref{Un}) take the forms
\begin{eqnarray}
{\hat H}_{eff}^{(1)} &=&  2\,\delta_1 s_1 \left(\frac{2}{3}{\hat P}^{(g_1)} -\frac{1}{3} \sin(2kz +\Delta\phi) \,\frac{{\hat F}}{F_{g_1}} \right)\,, \nonumber \\
{\hat H}_{eff}^{(2)} &=& 2\, \delta_2 s_2 \left(\frac{1}{3}{\hat P}^{(g_2)} -\frac{1}{9} \sin(2kz)\, \frac{{\hat F}}{F_{g_2}} \right) \, .
\end{eqnarray}
The force on an atom in this field can also be divided into the sum of two parts
\begin{widetext}
\begin{eqnarray}\label{fzlinplin}
f_1 &=& - \hbar k\frac{8}{3} \frac{\delta_1 s_1 s_2}{Q} \cos(2kz+2\Delta\phi) \left[ \sin(2kz +2\Delta\phi)\sin(2kz) +5 \right]\left[2\sin(2kz +2\Delta\phi)+\sin(2kz) \right] \,,\nonumber \\
f_2 &=& - \hbar k\frac{4}{3} \frac{\delta_2 s_1 s_2}{Q} \cos^2(2kz+2\Delta\phi) \cos(kz) \left[2\sin(2kz +2\Delta\phi)+\sin(2kz) \right]\,, \\
Q &=&s_1 9\cos^2(2kz+2\Delta\phi) +
s_2\left[20-\cos(4kz)-\cos(4kz+4\Delta\phi) +14 \sin(2kz)\sin(2kz+\Delta\phi)  \right] \,. \nonumber
\end{eqnarray}
\end{widetext}
As in the case above, the rectified force in double $lin$\,$\perp$\,$lin$ field configuration creates a macroscopic potential, that takes zero values at points where the relative phase of the fields is $\Delta\phi = 0, \pm \pi/2, \pm \pi \dots$. The macroscopic potential is determined by the average force over the wavelength (\ref{u12lin0}). For the double $lin$\,$\perp$\,$lin$ configuration, the analytical expression for the depths of the macroscopic potentials $\Delta U_n$ [see Eq.\,(\ref{DeltaU_n})]  are quite complicated. However, they can also be represented in the form (\ref{u1u2_lin0}), where the dependencies of the dimensionless functions $\mu_1$ and $\mu_2$ on the saturation parameter ratio $s_2/s_1$ are presented in Fig.\,\ref{fig:uu_LinPLin}.

\begin{figure}[h]
\centerline{\includegraphics[width=3.2 in]{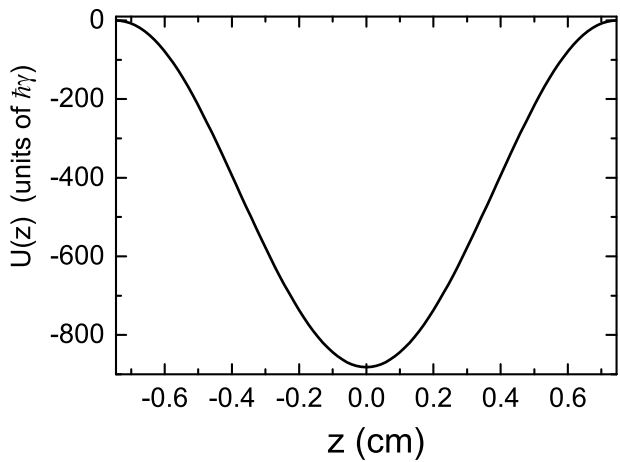}}
\caption{Macroscopic optical potential $U(z)$ in units of $\hbar
\gamma$ of a bichromatic lattice formed by field of double
$lin$\,$\perp$\,$lin$ configuration.  The detunings are $\delta_1 =
-1.5 \gamma$ and $\delta_2 = - \gamma$, the saturation parameters
are $s_1 = 0.1$, $s_2 = 0.3$.  The depth of macroscopic optical
potential for considered parameters reaches $\Delta U = 890\, \hbar
\gamma$ that corresponds $\simeq 0.25$ K in temperature units.}
\label{fig:uzlin2}
\end{figure}

\begin{figure}[h]
\centerline{\includegraphics[width=3.8 in]{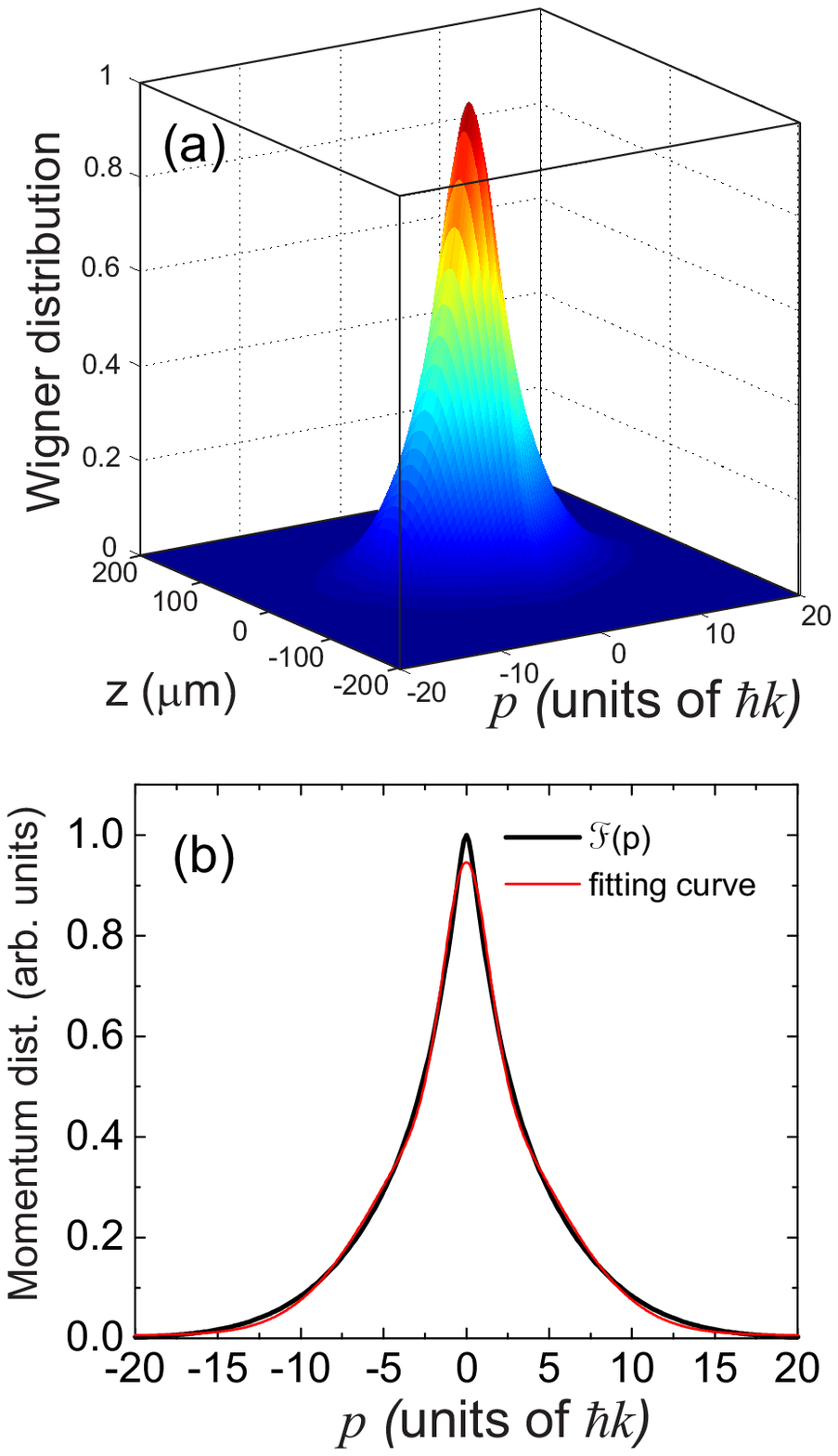}} \caption{The
Wigner function of the atomic phase distribution ${\cal F}(z,p)$ (a)
and momentum distribution of cold atoms (b) in the macroscopic
potential of the bichromatic field of double $lin$\,$\perp$\,$lin$
configuration. The waves parameters correspond to
Fig.\ref{fig:uzlin2}.} \label{fig:W2}
\end{figure}

The macroscopic potential for the double field $lin$\,$\perp$\,$lin$ is presented in Fig.\,\ref{fig:uzlin2}. The main feature of the double $lin$\,$\perp$\,$lin$ configuration is the presence of polarization gradient cooling mechanisms, which lead to the possibility of sub-Doppler cooling \cite{dal1989}. Indeed, for low field intensity we observe narrow distributions of trapped atoms in the phase space, as shown in Fig.\,\ref{fig:W2}. Note that in the conditions of sub-Doppler cooling, the distribution of atoms in the momentum space is essentially non-equilibrium and generally cannot be described in terms of temperature. However, as shown in \cite{Kalganova}, the momentum distribution of atoms is well approximated by a two-temperature Gaussian functions.
The momentum distribution of trapped atoms  presented in Fig.\,\ref{fig:W2}(b) is approximated by a two-temperature Gaussian function with temperature $T_{C} \simeq 0.035\, \hbar \gamma/k_B \simeq 10 \mu K$ for ``cold'' fraction of atoms $N_C \simeq 20 \%$, and  temperature  $T_{H} \simeq 0.5\, \hbar \gamma/k_B \simeq 140\, \mu K$ for ``hot'' fraction of atoms $N_H \simeq 80\%$.
For the parameters of Fig.\,\ref{fig:uzlin2} the size of trapped atoms cloud (\ref{sizeZ}) in the macroscopic potential is  $\beta_z \simeq 45\, \mu m$. Thus, the sub-Doppler cooling mechanisms  allow to significantly decrease the trapped atoms cloud size in comparison with double $lin\,||\,lin$ configuration.

\subsection{Multidimensional optical lattices created by a bichromatic field}
\begin{figure}[h]
\centerline{\includegraphics[width=3.5 in]{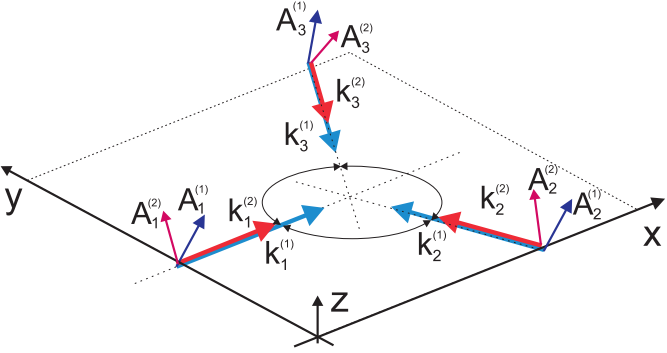}}
\caption{Two-dimensional configuration of bichromatic optical
lattice with phase independent topology.} \label{fig:field2D}
\end{figure}

In general, by combining bichromatic waves, as well as for monochromatic fields, one can create sufficiently complex spatial configurations of multidimensional optical lattices \cite{jessen96}. The spatial topology of light shifts (\ref{Un}) for each frequency component $\omega_1$ and $\omega_2$ depends not only on the chosen geometry of the waves and their polarizations, but may also depend on the relative phases of the light waves forming the field of each frequency component separately.
The phase dependence of the topology of optical lattices poses certain challenges for their experimental implementation and requires some efforts to control and stabilize the phase of the light waves forming the field \cite{Hemmerich1993}. Nevertheless, among all the diversity, it is possible to distinguish phase-independent configurations of optical lattices. Thus, in the case where the number of waves $M$ forming the field for each of the frequency components exceeds the dimension of the space $N$ by 1, $M=N+1$, then the light shifts formed by each frequency component (\ref{Un}) have a phase-independent topology \cite {jessen96}. We also note that the processes of optical pumping between the ground-state levels $|g_1 \rangle$  and $|g_2 \rangle$ are determined by quadratic combinations of the amplitudes of the light waves of each frequency component individually (\ref{reduced}) to (\ref{reducedeee}), and are not dependent on the relative temporal phases between different frequency components of the bichromatic field.
Thus, for the field configuration  leading to the phase-independent topology of monochromatic lattices, the bichromatic field configuration will also lead to a phase-independent topology for a bichromatic lattice with a macroscopic period for lithium atoms.
\begin{figure}[h]
\centerline{\includegraphics[width=3.2 in]{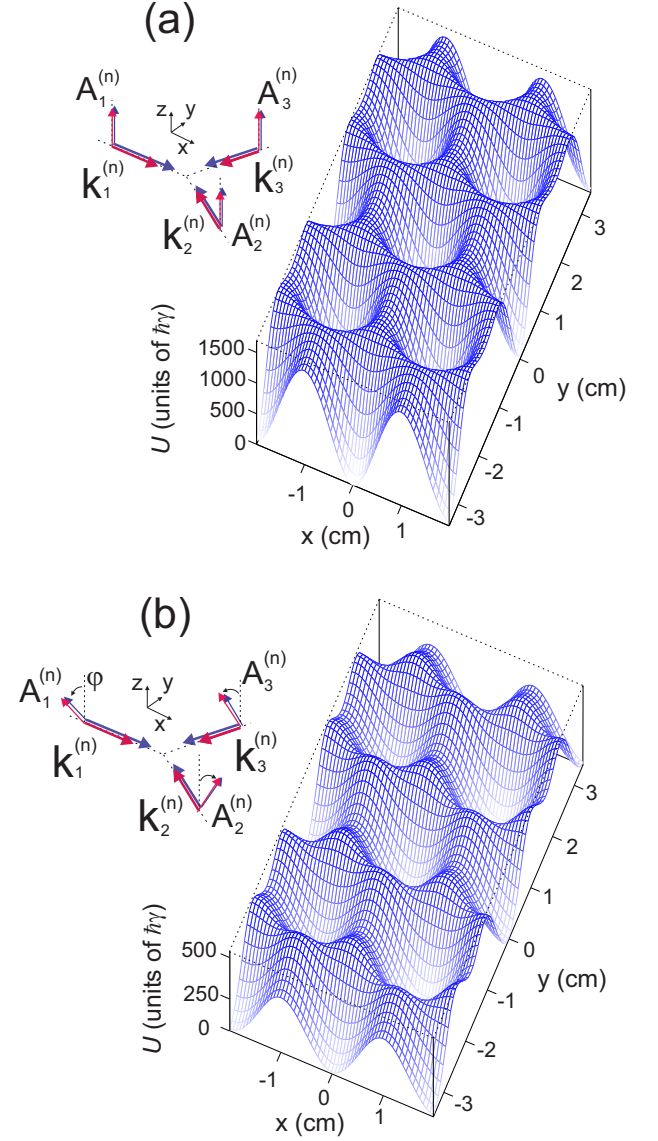}} \caption{(a)
Macroscopic $2$D optical potential formed by a bichromatic field of
light waves with linear polarization along the $z$ axis. The
parameters of the waves correspond to the parameters of Figs.
\ref{fig:uzlin0}. (b) Macroscopic $2$D optical potential generated
by a bichromatic field of linearly polarized light waves oriented at
an angle $\varphi = \pi/4$ to the $z$ axis. The parameters of the
waves correspond to the parameters of Fig. \ref{fig:uzlin2}.}
\label{fig:U2D}
\end{figure}

An example of a two-dimensional (2D) configuration of fields forming a phase-independent bichromatic optical lattice is presented in Fig.\,\ref{fig:field2D}, where the angles between the wave vectors of the running waves are 120$^{\circ}$.  Figure\,\ref{fig:U2D} shows the spatial dependencies of the optical potentials formed by three pairs of waves in a symmetric configuration of Fig.\,\ref{fig:field2D}. There are two cases represented: the waves with linear polarization along the $z$ axis [see Fig.\,\ref{fig:U2D}(a)], and the waves with linear polarizations oriented at an angle $\varphi = \pi/4$ to the $z$ axis  [see Fig.\,\ref{fig:U2D}(b)]. In the second case, the field, in addition to spatial non-uniform intensity, also has spatial non-uniformity of the local polarization parameters (ellipticity and orientation) of the fields ${\bf E}^{(1)}({\bf r})$ and ${\bf E}^{(2) }({\bf r})$. As is well known, the spatial non-uniformity of polarization leads to sub-Doppler mechanisms of laser cooling \cite{dal1993EPL,pru2007j}.
The periods of macroscopic potential formed by three pairs of light waves in Fig.\,\ref{fig:field2D} along the $x$ and $y$ axes are
\begin{eqnarray}
\Lambda_x &=& \frac{\lambda}{3}\frac{k}{\Delta k} \simeq 1.9 \, \,\mbox{cm}\,, \nonumber \\
\Lambda_y &=& \frac{\lambda}{\sqrt{3}}\frac{k}{\Delta k} \simeq 3.4 \,\, \mbox{cm} \, .
\end{eqnarray}
Note that the depths of the bichromatic potentials Fig.\,\ref{fig:U2D}(a) and (b) in order of magnitude correspond to the potential depths for the one-dimensional configurations Fig.\,\ref{fig:uzlin0} and Fig.\,\ref{fig:uzlin2} respectively.

\begin{figure}[t]
\centerline{\includegraphics[width=3.4 in]{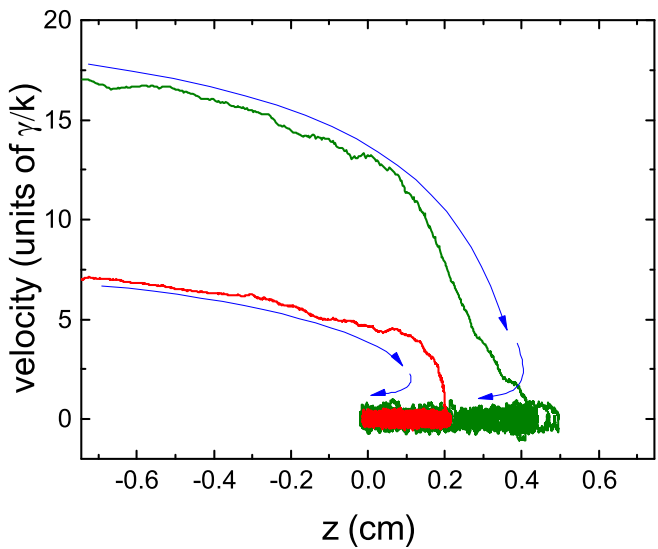}} \caption{The
phase-space trajectory of an atom entering to the macroscopic trap
formed by double $lin \perp lin$ configuration of bichromatic field.
The red (lower) line corresponds to atom trajectory in the trap with
parameters of Fig. \ref{fig:uzlin2} ($\delta_1 = -1.5 \gamma$,
$\delta_2 = - \gamma$, $s_1 = 0.1$, $s_2 = 0.3$). The initial
velocity of atoms on the trap boundary $v = 7\, \gamma/k\simeq 25$
m/s. The green (upper) line corresponds to the atom's trajectory in
the trap with $\delta_1 = -10 \gamma$, $\delta_2 = -2.5 \gamma$,
$s_1 = 0.05$, $s_2 = 0.5$. The initial velocity of atoms on the trap
boundary $v = 17\, \gamma/k\simeq 60$ m/s. The blue arrows define
the direction of the trajectory evolution.   } \label{fig:13}
\end{figure}

An alternative way to avoid phase dependence in the  topology of the multidimensional optical trap formed by a combination of bichromatic standing waves is to use the waves along different directions with slightly shifted frequencies. For the frequency shift of a few MHz the mutual phase gets enough fast oscillation and is much faster then the rate of the kinetic evolution processes.
In this case the atom's motion will be determined by the phase-averaged dissipative force that also forms 2D, or three-dimensional (3D) optical traps with macroscopic period.

The presented deep macroscopic potentials can be used as an alternative to magneto-optical traps for the cooling and  simultaneous trapping of lithium atoms. The number of atoms trapped from vapors can be estimated using the relation presented in \cite{Monroe1990,Chu1992}
\begin{equation}\label{Ncc}
    N_{c}=\frac{L^2}{\sigma_c}\left(\frac{v_c}{u}\right)^4 \,,
\end{equation}
where $v_c$ is the capture velocity, $u = (2 k_B T/m)^{1/2}$ is the most probable speed, $L$ is the size of the trap determined by the size of the laser beam, and $\sigma_c$ is the cross section for an atom to eject atom from a trap. The capture velocity can be found from the analysis of trajectories of atoms coming to the trap with different velocities, separating those that lead to the capturing of atoms. This analysis can be done on the basis of Langevin equations (see, for example, \cite{Langevin,Langevin2}) describing the atom trajectory under the light force and associated stochastic diffusion. Here we used numerical expressions for the force and diffusion as functions of velocity and the position of the atom.  Figure\,\ref{fig:13} demonstrates the trajectory in the phase space of the atom captured in the macroscopic potential formed by bichromatic field of double $lin \perp lin$ configuration.
The direct numerical analysis results in $v_c \simeq 10\, k/\gamma \simeq 35$ m/s for the traps with parameters of  Fig.\,\ref{fig:uzlin2}, which leads to estimation $N_c \simeq 10^7$ for the number of trapped atoms. We additionally note that the number of trapped atoms $N_c$ is determined not only by the capture velocity, $v_c$, which can be increased by adjusting the parameters of the bichromatic field, but also by the most probable velocity at which atoms enter the trap, $u$ (\ref{Ncc}). For lithium atoms loaded from vapors $u \simeq 900$ m/s (at temperature T=300K) is sufficiently large compared to heavy elements ($190$ m/s for Cs atoms and $240$ m/s for Rb atoms at temperature T = 300\,K).
Therefore, in many works, the methods of pre-cooling and slowing down atoms are used to increase the number of lithium atoms in MOT \cite{SCHUNEMANN1998263,Randall2020,Rajkov2020}. A similar technique can also be used to increase the number of atoms loaded into a bichromatic trap.

\section{Conclusion}
In this paper, we present an idea of a deep pure-optical trap for neutral atoms based on a dissipative bichromatic optical lattice. For lithium atoms, as an example, a low-intensity bichromatic field (about 5\,  mW/cm$^2$) allows for the creation of a deep optical potential with a macroscopic period of centimeter scale, which provides laser cooling and trapping of atoms. The analysis, which was carried out within the one-atom approximation, showed that the temperature of laser cooling can reach values comparable to and below the Doppler limit (is approximately $\simeq 140\, \mu K$ for lithium atoms).  At the same time, the localization of trapped atoms reaches sub-millimeter sizes, and the number of trapped atoms is comparable to the number trapped in MOT.

It should also be noted that the presented idea of deep dissipative lattice with a macroscopic period in a bichromatic field can be generalized to other neutral atoms. For example, for Rb and Cs atoms transitions between hyperfine components of $D_2$ line can be used as resonant transitions for two frequency components of bichromatic lattice. In this case, the period of the macroscopic potential will be determined mainly by the splitting between the hyperfine components of the ground state $^2S_{1/2}$ and gives: $\Lambda \simeq  4.9$ cm for $^{85}$Rb, $\Lambda \simeq 2.2$ cm for $^{87}$Rb, and $\Lambda \simeq  1.6$ cm for $^{133}$Cs. Also, for odd isotopes of mercury atoms $^{199}$Hg and $^{201}$Hg, the intercombination transition $^1S_0 \to \, ^3P_1$ ($\lambda = 253.7$ nm, $\gamma/ 2\pi = 1.3$ MHz) can be used, where hyperfine splitting of the upper level $^3P_1$ exceeds 20 GHz that allows to create a bichromatic lattice with  macroscopic period of $\Lambda \simeq 0.7$ cm.

Such deep pure-optical macroscopic potential can be used for cooling and trapping of atoms in compact devices with low power consumption.

We thank Ennio Arimondo for fruitful discussion and valuable comments.

The research was supported by the Russian Science Foundation (Project No 23-22-00198),
https://rscf.ru/project/23-22-00198/

\nocite{*}

\bibliography{paper}

\begin{thebibliography}{45}
\expandafter\ifx\csname natexlab\endcsname\relax\def\natexlab#1{#1}\fi
\expandafter\ifx\csname bibnamefont\endcsname\relax
  \def\bibnamefont#1{#1}\fi
\expandafter\ifx\csname bibfnamefont\endcsname\relax
  \def\bibfnamefont#1{#1}\fi
\expandafter\ifx\csname citenamefont\endcsname\relax
  \def\citenamefont#1{#1}\fi
\expandafter\ifx\csname url\endcsname\relax
  \def\url#1{\texttt{#1}}\fi
\expandafter\ifx\csname urlprefix\endcsname\relax\def\urlprefix{URL }\fi
\providecommand{\bibinfo}[2]{#2}
\providecommand{\eprint}[2][]{\url{#2}}

\bibitem[{\citenamefont{Raab et~al.}(1987)\citenamefont{Raab, Prentis, Cable,
  Chu, and Pritchard}}]{raab1987}
\bibinfo{author}{\bibfnamefont{E.}~\bibnamefont{Raab}},
  \bibinfo{author}{\bibfnamefont{M.}~\bibnamefont{Prentis}},
  \bibinfo{author}{\bibfnamefont{A.}~\bibnamefont{Cable}},
  \bibinfo{author}{\bibfnamefont{S.}~\bibnamefont{Chu}}, \bibnamefont{and}
  \bibinfo{author}{\bibfnamefont{D.}~\bibnamefont{Pritchard}},
  \bibinfo{journal}{Phys. Lev. Lett.} \textbf{\bibinfo{volume}{59}},
  \bibinfo{pages}{2631} (\bibinfo{year}{1987}).

\bibitem[{\citenamefont{Metcalf and Straten}(1999)}]{metcalf}
\bibinfo{author}{\bibfnamefont{H.~J.} \bibnamefont{Metcalf}} \bibnamefont{and}
  \bibinfo{author}{\bibfnamefont{P.~v.~d.} \bibnamefont{Straten}},
  \emph{\bibinfo{title}{Laser Cooling and Trapping}}
  (\bibinfo{publisher}{Springer}, \bibinfo{year}{1999}).

\bibitem[{\citenamefont{Berman}(1997)}]{berman}
\bibinfo{author}{\bibfnamefont{P.~R.} \bibnamefont{Berman}},
  \emph{\bibinfo{title}{Atom Interferometry}} (\bibinfo{publisher}{Academic
  Press}, \bibinfo{address}{San Diego}, \bibinfo{year}{1997}), ISBN
  \bibinfo{isbn}{978-0-12-092460-8}.

\bibitem[{\citenamefont{Takamoto et~al.}(2020)\citenamefont{Takamoto, Ushijima,
  Ohmae, Yahagi, Kokado, Shinkai, and Katori}}]{Katori2020}
\bibinfo{author}{\bibfnamefont{M.}~\bibnamefont{Takamoto}},
  \bibinfo{author}{\bibfnamefont{I.}~\bibnamefont{Ushijima}},
  \bibinfo{author}{\bibfnamefont{N.}~\bibnamefont{Ohmae}},
  \bibinfo{author}{\bibfnamefont{T.}~\bibnamefont{Yahagi}},
  \bibinfo{author}{\bibfnamefont{K.}~\bibnamefont{Kokado}},
  \bibinfo{author}{\bibfnamefont{H.}~\bibnamefont{Shinkai}}, \bibnamefont{and}
  \bibinfo{author}{\bibfnamefont{H.}~\bibnamefont{Katori}},
  \bibinfo{journal}{Nat. Photonics} \textbf{\bibinfo{volume}{14}},
  \bibinfo{pages}{411–415} (\bibinfo{year}{2020}).

\bibitem[{\citenamefont{Lion et~al.}(2017)\citenamefont{Lion, Panet, Wolf,
  Guerlin, Bize, and Delva}}]{Lion2017}
\bibinfo{author}{\bibfnamefont{G.}~\bibnamefont{Lion}},
  \bibinfo{author}{\bibfnamefont{I.}~\bibnamefont{Panet}},
  \bibinfo{author}{\bibfnamefont{P.}~\bibnamefont{Wolf}},
  \bibinfo{author}{\bibfnamefont{C.}~\bibnamefont{Guerlin}},
  \bibinfo{author}{\bibfnamefont{S.}~\bibnamefont{Bize}}, \bibnamefont{and}
  \bibinfo{author}{\bibfnamefont{P.}~\bibnamefont{Delva}},
  \bibinfo{journal}{Journal of Geodesy} \textbf{\bibinfo{volume}{91}},
  \bibinfo{pages}{597–611} (\bibinfo{year}{2017}).

\bibitem[{\citenamefont{McGrew et~al.}(2018)\citenamefont{McGrew, Zhang,
  Fasano, Schäffer, Beloy, Nicolodi, Brown, Hinkley, Milani, Schioppo
  et~al.}}]{Ludlow2018}
\bibinfo{author}{\bibfnamefont{W.~F.} \bibnamefont{McGrew}},
  \bibinfo{author}{\bibfnamefont{X.}~\bibnamefont{Zhang}},
  \bibinfo{author}{\bibfnamefont{R.~J.} \bibnamefont{Fasano}},
  \bibinfo{author}{\bibfnamefont{S.~A.} \bibnamefont{Schäffer}},
  \bibinfo{author}{\bibfnamefont{K.}~\bibnamefont{Beloy}},
  \bibinfo{author}{\bibfnamefont{D.}~\bibnamefont{Nicolodi}},
  \bibinfo{author}{\bibfnamefont{R.~C.} \bibnamefont{Brown}},
  \bibinfo{author}{\bibfnamefont{N.}~\bibnamefont{Hinkley}},
  \bibinfo{author}{\bibfnamefont{G.}~\bibnamefont{Milani}},
  \bibinfo{author}{\bibfnamefont{M.}~\bibnamefont{Schioppo}},
  \bibnamefont{et~al.}, \bibinfo{journal}{Nature}
  \textbf{\bibinfo{volume}{564}}, \bibinfo{pages}{87} (\bibinfo{year}{2018}).

\bibitem[{\citenamefont{Nicholson et~al.}(2015)\citenamefont{Nicholson,
  Campbell, Hutson, Bloom, McNally, Zhang, Barret, Safronova, Strouse, Tew
  et~al.}}]{Nicholson}
\bibinfo{author}{\bibfnamefont{T.}~\bibnamefont{Nicholson}},
  \bibinfo{author}{\bibfnamefont{S.}~\bibnamefont{Campbell}},
  \bibinfo{author}{\bibfnamefont{R.}~\bibnamefont{Hutson}},
  \bibinfo{author}{\bibfnamefont{B.}~\bibnamefont{Bloom}},
  \bibinfo{author}{\bibfnamefont{R.}~\bibnamefont{McNally}},
  \bibinfo{author}{\bibfnamefont{W.}~\bibnamefont{Zhang}},
  \bibinfo{author}{\bibfnamefont{M.}~\bibnamefont{Barret}},
  \bibinfo{author}{\bibfnamefont{M.}~\bibnamefont{Safronova}},
  \bibinfo{author}{\bibfnamefont{G.}~\bibnamefont{Strouse}},
  \bibinfo{author}{\bibfnamefont{W.}~\bibnamefont{Tew}}, \bibnamefont{et~al.},
  \bibinfo{journal}{Nat. Commun.} \textbf{\bibinfo{volume}{6}},
  \bibinfo{pages}{6896} (\bibinfo{year}{2015}).

\bibitem[{\citenamefont{Jessen and Deutsch}(1996)}]{jessen96}
\bibinfo{author}{\bibfnamefont{P.}~\bibnamefont{Jessen}} \bibnamefont{and}
  \bibinfo{author}{\bibfnamefont{I.}~\bibnamefont{Deutsch}},
  \bibinfo{journal}{Advances In Atomic, Molecular, and Optical Physics}
  \textbf{\bibinfo{volume}{37}}, \bibinfo{pages}{95} (\bibinfo{year}{1996}).

\bibitem[{\citenamefont{Grynberg and Robilliard}(2001)}]{Grynberg}
\bibinfo{author}{\bibfnamefont{G.}~\bibnamefont{Grynberg}} \bibnamefont{and}
  \bibinfo{author}{\bibfnamefont{C.}~\bibnamefont{Robilliard}},
  \bibinfo{journal}{Physics Reports} \textbf{\bibinfo{volume}{355}},
  \bibinfo{pages}{335–451} (\bibinfo{year}{2001}).

\bibitem[{\citenamefont{Dalibard and Cohen-Tannoudji}(1989)}]{dal1989}
\bibinfo{author}{\bibfnamefont{J.}~\bibnamefont{Dalibard}} \bibnamefont{and}
  \bibinfo{author}{\bibfnamefont{C.}~\bibnamefont{Cohen-Tannoudji}},
  \bibinfo{journal}{J. Opt. Soc. Am. B} \textbf{\bibinfo{volume}{6}},
  \bibinfo{pages}{2023} (\bibinfo{year}{1989}).

\bibitem[{\citenamefont{Berg-S{\o}rensen
  et~al.}(1993)\citenamefont{Berg-S{\o}rensen, Castin, M{\o}lmer, and
  Dalibard}}]{dal1993EPL}
\bibinfo{author}{\bibfnamefont{K.}~\bibnamefont{Berg-S{\o}rensen}},
  \bibinfo{author}{\bibfnamefont{Y.}~\bibnamefont{Castin}},
  \bibinfo{author}{\bibfnamefont{K.}~\bibnamefont{M{\o}lmer}},
  \bibnamefont{and} \bibinfo{author}{\bibfnamefont{J.}~\bibnamefont{Dalibard}},
  \bibinfo{journal}{Europhysics Letters} \textbf{\bibinfo{volume}{22}},
  \bibinfo{pages}{663} (\bibinfo{year}{1993}).

\bibitem[{\citenamefont{Prudnikov
  et~al.}(2007{\natexlab{a}})\citenamefont{Prudnikov, Taichenachev, Tumaikin,
  and Yudin}}]{pru2007j}
\bibinfo{author}{\bibfnamefont{O.~N.} \bibnamefont{Prudnikov}},
  \bibinfo{author}{\bibfnamefont{A.~V.} \bibnamefont{Taichenachev}},
  \bibinfo{author}{\bibfnamefont{A.~V.} \bibnamefont{Tumaikin}},
  \bibnamefont{and} \bibinfo{author}{\bibfnamefont{V.~I.} \bibnamefont{Yudin}},
  \bibinfo{journal}{JETP} \textbf{\bibinfo{volume}{104}}, \bibinfo{pages}{839}
  (\bibinfo{year}{2007}{\natexlab{a}}).

\bibitem[{\citenamefont{Prudnikov and Arimondo}(2004{\natexlab{a}})}]{pru2004}
\bibinfo{author}{\bibfnamefont{O.~N.} \bibnamefont{Prudnikov}}
  \bibnamefont{and} \bibinfo{author}{\bibfnamefont{E.}~\bibnamefont{Arimondo}},
  \bibinfo{journal}{Journal of Optics B: Quantum and Semiclassical Optics}
  \textbf{\bibinfo{volume}{6}}, \bibinfo{pages}{336}
  (\bibinfo{year}{2004}{\natexlab{a}}).

\bibitem[{\citenamefont{Dalibard and
  Cohen-Tannoudji}(1985{\natexlab{a}})}]{Dal85d}
\bibinfo{author}{\bibfnamefont{J.}~\bibnamefont{Dalibard}} \bibnamefont{and}
  \bibinfo{author}{\bibfnamefont{C.}~\bibnamefont{Cohen-Tannoudji}},
  \bibinfo{journal}{J. Opt. Soc. Am. B} \textbf{\bibinfo{volume}{2}},
  \bibinfo{pages}{1707} (\bibinfo{year}{1985}{\natexlab{a}}).

\bibitem[{\citenamefont{Prudnikov
  et~al.}(2007{\natexlab{b}})\citenamefont{Prudnikov, Taichenachev, Tumaikin,
  and I.}}]{pru2007}
\bibinfo{author}{\bibfnamefont{O.~N.} \bibnamefont{Prudnikov}},
  \bibinfo{author}{\bibfnamefont{A.~V.} \bibnamefont{Taichenachev}},
  \bibinfo{author}{\bibfnamefont{A.~M.} \bibnamefont{Tumaikin}},
  \bibnamefont{and} \bibinfo{author}{\bibfnamefont{Y.~V.} \bibnamefont{I.}},
  \bibinfo{journal}{Phys. Rev. A} \textbf{\bibinfo{volume}{75}},
  \bibinfo{pages}{023413} (\bibinfo{year}{2007}{\natexlab{b}}).

\bibitem[{\citenamefont{Prudnikov et~al.}(2011)\citenamefont{Prudnikov,
  Ilenkov, Taichenachev, Tumaikin, and Yudin}}]{pru2011}
\bibinfo{author}{\bibfnamefont{O.~N.} \bibnamefont{Prudnikov}},
  \bibinfo{author}{\bibfnamefont{R.~Y.} \bibnamefont{Ilenkov}},
  \bibinfo{author}{\bibfnamefont{A.~V.} \bibnamefont{Taichenachev}},
  \bibinfo{author}{\bibfnamefont{A.~M.} \bibnamefont{Tumaikin}},
  \bibnamefont{and} \bibinfo{author}{\bibfnamefont{V.~I.} \bibnamefont{Yudin}},
  \bibinfo{journal}{JETP} \textbf{\bibinfo{volume}{112}}, \bibinfo{pages}{939}
  (\bibinfo{year}{2011}).

\bibitem[{\citenamefont{Kulosa et~al.}(2015)\citenamefont{Kulosa, Fim, Zipfel,
  R\"uhmann, Sauer, Jha, Gibble, Ertmer, Rasel, Safronova et~al.}}]{Rassel2015}
\bibinfo{author}{\bibfnamefont{A.~P.} \bibnamefont{Kulosa}},
  \bibinfo{author}{\bibfnamefont{D.}~\bibnamefont{Fim}},
  \bibinfo{author}{\bibfnamefont{K.~H.} \bibnamefont{Zipfel}},
  \bibinfo{author}{\bibfnamefont{S.}~\bibnamefont{R\"uhmann}},
  \bibinfo{author}{\bibfnamefont{S.}~\bibnamefont{Sauer}},
  \bibinfo{author}{\bibfnamefont{N.}~\bibnamefont{Jha}},
  \bibinfo{author}{\bibfnamefont{K.}~\bibnamefont{Gibble}},
  \bibinfo{author}{\bibfnamefont{W.}~\bibnamefont{Ertmer}},
  \bibinfo{author}{\bibfnamefont{E.~M.} \bibnamefont{Rasel}},
  \bibinfo{author}{\bibfnamefont{M.~S.} \bibnamefont{Safronova}},
  \bibnamefont{et~al.}, \bibinfo{journal}{Phys. Rev. Lett.}
  \textbf{\bibinfo{volume}{115}}, \bibinfo{pages}{240801}
  (\bibinfo{year}{2015}).

\bibitem[{\citenamefont{Kazantsev and Krasnov}(1987)}]{kaz87}
\bibinfo{author}{\bibfnamefont{A.~P.} \bibnamefont{Kazantsev}}
  \bibnamefont{and} \bibinfo{author}{\bibfnamefont{I.~V.}
  \bibnamefont{Krasnov}}, \bibinfo{journal}{JETP Lett.}
  \textbf{\bibinfo{volume}{46}}, \bibinfo{pages}{332} (\bibinfo{year}{1987}).

\bibitem[{\citenamefont{Voitsekhovich et~al.}(1988)\citenamefont{Voitsekhovich,
  Danileiko, Negriiko, Romanenko, and P.}}]{Voitsekhovich88}
\bibinfo{author}{\bibfnamefont{V.~S.} \bibnamefont{Voitsekhovich}},
  \bibinfo{author}{\bibfnamefont{M.~V.} \bibnamefont{Danileiko}},
  \bibinfo{author}{\bibfnamefont{A.~M.} \bibnamefont{Negriiko}},
  \bibinfo{author}{\bibfnamefont{V.~I.} \bibnamefont{Romanenko}},
  \bibnamefont{and} \bibinfo{author}{\bibfnamefont{Y.~L.} \bibnamefont{P.}},
  \bibinfo{journal}{Zhurnal Tekhnicheskoi Fiziki}
  \textbf{\bibinfo{volume}{58}}, \bibinfo{pages}{1174–1176}
  (\bibinfo{year}{1988}).

\bibitem[{\citenamefont{Grimm et~al.}(1990)\citenamefont{Grimm, Ovchinnikov,
  Sidorov, and Letokhov}}]{ovch90}
\bibinfo{author}{\bibfnamefont{R.}~\bibnamefont{Grimm}},
  \bibinfo{author}{\bibfnamefont{Y.~B.} \bibnamefont{Ovchinnikov}},
  \bibinfo{author}{\bibfnamefont{A.~I.} \bibnamefont{Sidorov}},
  \bibnamefont{and} \bibinfo{author}{\bibfnamefont{V.~S.}
  \bibnamefont{Letokhov}}, \bibinfo{journal}{Phys. Rev. Lett.}
  \textbf{\bibinfo{volume}{65}}, \bibinfo{pages}{1415} (\bibinfo{year}{1990}).

\bibitem[{\citenamefont{S\"oding et~al.}(1997)\citenamefont{S\"oding, Grimm,
  Ovchinnikov, Bouyer, and Salomon}}]{Soding97}
\bibinfo{author}{\bibfnamefont{J.}~\bibnamefont{S\"oding}},
  \bibinfo{author}{\bibfnamefont{R.}~\bibnamefont{Grimm}},
  \bibinfo{author}{\bibfnamefont{Y.~B.} \bibnamefont{Ovchinnikov}},
  \bibinfo{author}{\bibfnamefont{P.}~\bibnamefont{Bouyer}}, \bibnamefont{and}
  \bibinfo{author}{\bibfnamefont{C.}~\bibnamefont{Salomon}},
  \bibinfo{journal}{Phys. Rev. Lett.} \textbf{\bibinfo{volume}{78}},
  \bibinfo{pages}{1420} (\bibinfo{year}{1997}).

\bibitem[{\citenamefont{Cashen and Metcalf}(2001)}]{Metcalf_He}
\bibinfo{author}{\bibfnamefont{M.~T.} \bibnamefont{Cashen}} \bibnamefont{and}
  \bibinfo{author}{\bibfnamefont{H.}~\bibnamefont{Metcalf}},
  \bibinfo{journal}{Phys. Rev. A} \textbf{\bibinfo{volume}{63}},
  \bibinfo{pages}{025406} (\bibinfo{year}{2001}).

\bibitem[{\citenamefont{Liebisch et~al.}(2012)\citenamefont{Liebisch, Blanshan,
  Donley, and Kitching}}]{Kitching2012}
\bibinfo{author}{\bibfnamefont{T.~C.} \bibnamefont{Liebisch}},
  \bibinfo{author}{\bibfnamefont{E.}~\bibnamefont{Blanshan}},
  \bibinfo{author}{\bibfnamefont{E.~A.} \bibnamefont{Donley}},
  \bibnamefont{and} \bibinfo{author}{\bibfnamefont{J.}~\bibnamefont{Kitching}},
  \bibinfo{journal}{Phys. Rev. A} \textbf{\bibinfo{volume}{85}},
  \bibinfo{pages}{013407} (\bibinfo{year}{2012}).

\bibitem[{\citenamefont{Prudnikov et~al.}(2013)\citenamefont{Prudnikov,
  Baklanov, Taichenachev, Tumaikin, and Yudin}}]{pru2013}
\bibinfo{author}{\bibfnamefont{O.~N.} \bibnamefont{Prudnikov}},
  \bibinfo{author}{\bibfnamefont{A.~S.} \bibnamefont{Baklanov}},
  \bibinfo{author}{\bibfnamefont{A.~V.} \bibnamefont{Taichenachev}},
  \bibinfo{author}{\bibfnamefont{A.~M.} \bibnamefont{Tumaikin}},
  \bibnamefont{and} \bibinfo{author}{\bibfnamefont{V.~I.} \bibnamefont{Yudin}},
  \bibinfo{journal}{JETP} \textbf{\bibinfo{volume}{117}}, \bibinfo{pages}{222}
  (\bibinfo{year}{2013}).

\bibitem[{\citenamefont{Prudnikov et~al.}(2017)\citenamefont{Prudnikov,
  Taichenachev, and Yudin}}]{pru2017}
\bibinfo{author}{\bibfnamefont{O.~N.} \bibnamefont{Prudnikov}},
  \bibinfo{author}{\bibfnamefont{A.~V.} \bibnamefont{Taichenachev}},
  \bibnamefont{and} \bibinfo{author}{\bibfnamefont{V.~I.} \bibnamefont{Yudin}},
  \bibinfo{journal}{Quantum Electronics} \textbf{\bibinfo{volume}{47}},
  \bibinfo{pages}{438–445} (\bibinfo{year}{2017}).

\bibitem[{\citenamefont{Khersonskii et~al.}(1988)\citenamefont{Khersonskii,
  Moskalev, and Varshalovich}}]{Varshalovich}
\bibinfo{author}{\bibfnamefont{V.~K.} \bibnamefont{Khersonskii}},
  \bibinfo{author}{\bibfnamefont{A.~N.} \bibnamefont{Moskalev}},
  \bibnamefont{and} \bibinfo{author}{\bibfnamefont{D.~A.}
  \bibnamefont{Varshalovich}}, \emph{\bibinfo{title}{Quantum Theory Of Angular
  Momentum}} (\bibinfo{publisher}{World Scientific Publishing Company,
  Singapore}, \bibinfo{year}{1988}).

\bibitem[{\citenamefont{Kirpichnikova et~al.}(2022)\citenamefont{Kirpichnikova,
  Prudnikov, Taichenachev, and Yudin}}]{kirp2022}
\bibinfo{author}{\bibfnamefont{A.~A.} \bibnamefont{Kirpichnikova}},
  \bibinfo{author}{\bibfnamefont{O.~N.} \bibnamefont{Prudnikov}},
  \bibinfo{author}{\bibfnamefont{A.~V.} \bibnamefont{Taichenachev}},
  \bibnamefont{and} \bibinfo{author}{\bibfnamefont{V.~I.} \bibnamefont{Yudin}},
  \bibinfo{journal}{Quantum Electronics} \textbf{\bibinfo{volume}{52}},
  \bibinfo{pages}{130 – 136} (\bibinfo{year}{2022}).

\bibitem[{\citenamefont{Kirpichnikova et~al.}(2020)\citenamefont{Kirpichnikova,
  Prudnikov, Il'enkov, Taichenachev, and Yudin}}]{kirp2020}
\bibinfo{author}{\bibfnamefont{A.~A.} \bibnamefont{Kirpichnikova}},
  \bibinfo{author}{\bibfnamefont{O.~N.} \bibnamefont{Prudnikov}},
  \bibinfo{author}{\bibfnamefont{R.~Y.} \bibnamefont{Il'enkov}},
  \bibinfo{author}{\bibfnamefont{A.~V.} \bibnamefont{Taichenachev}},
  \bibnamefont{and} \bibinfo{author}{\bibfnamefont{V.~I.} \bibnamefont{Yudin}},
  \bibinfo{journal}{Quantum Electronics} \textbf{\bibinfo{volume}{50}},
  \bibinfo{pages}{939 – 946} (\bibinfo{year}{2020}).

\bibitem[{\citenamefont{Dalibard and
  Cohen-Tannoudji}(1985{\natexlab{b}})}]{dal85sem}
\bibinfo{author}{\bibfnamefont{J.}~\bibnamefont{Dalibard}} \bibnamefont{and}
  \bibinfo{author}{\bibfnamefont{C.}~\bibnamefont{Cohen-Tannoudji}},
  \bibinfo{journal}{J.Phys. B.} \textbf{\bibinfo{volume}{18}},
  \bibinfo{pages}{1661} (\bibinfo{year}{1985}{\natexlab{b}}).

\bibitem[{\citenamefont{Javanainen}(1991)}]{Javanainen91}
\bibinfo{author}{\bibfnamefont{J.}~\bibnamefont{Javanainen}},
  \bibinfo{journal}{Phys. Rev. A} \textbf{\bibinfo{volume}{44}},
  \bibinfo{pages}{5857} (\bibinfo{year}{1991}).

\bibitem[{\citenamefont{Prudnikov
  et~al.}(1999{\natexlab{a}})\citenamefont{Prudnikov, Taichenachev, Tumaikin,
  and Yudin}}]{pru99}
\bibinfo{author}{\bibfnamefont{O.~N.} \bibnamefont{Prudnikov}},
  \bibinfo{author}{\bibfnamefont{A.~V.} \bibnamefont{Taichenachev}},
  \bibinfo{author}{\bibfnamefont{A.~M.} \bibnamefont{Tumaikin}},
  \bibnamefont{and} \bibinfo{author}{\bibfnamefont{V.~I.} \bibnamefont{Yudin}},
  \bibinfo{journal}{JETP} \textbf{\bibinfo{volume}{88}}, \bibinfo{pages}{433}
  (\bibinfo{year}{1999}{\natexlab{a}}).

\bibitem[{\citenamefont{Bezverbnyi et~al.}(2005)\citenamefont{Bezverbnyi,
  Prudnikov, Taichenachev, Tumaikin, and I.}}]{bez2005}
\bibinfo{author}{\bibfnamefont{A.~V.} \bibnamefont{Bezverbnyi}},
  \bibinfo{author}{\bibfnamefont{O.~N.} \bibnamefont{Prudnikov}},
  \bibinfo{author}{\bibfnamefont{A.~V.} \bibnamefont{Taichenachev}},
  \bibinfo{author}{\bibfnamefont{A.~V.} \bibnamefont{Tumaikin}},
  \bibnamefont{and} \bibinfo{author}{\bibfnamefont{Y.~V.} \bibnamefont{I.}},
  \bibinfo{journal}{JETP} \textbf{\bibinfo{volume}{101}}, \bibinfo{pages}{584}
  (\bibinfo{year}{2005}).

\bibitem[{\citenamefont{Hamann et~al.}(1998)\citenamefont{Hamann, Haycock,
  Klose, Pax, Deutsch, and S.}}]{jessen98}
\bibinfo{author}{\bibfnamefont{S.~E.} \bibnamefont{Hamann}},
  \bibinfo{author}{\bibfnamefont{D.~L.} \bibnamefont{Haycock}},
  \bibinfo{author}{\bibfnamefont{G.}~\bibnamefont{Klose}},
  \bibinfo{author}{\bibfnamefont{P.~H.} \bibnamefont{Pax}},
  \bibinfo{author}{\bibfnamefont{I.~H.} \bibnamefont{Deutsch}},
  \bibnamefont{and} \bibinfo{author}{\bibfnamefont{J.~P.} \bibnamefont{S.}},
  \bibinfo{journal}{PRL} \textbf{\bibinfo{volume}{80}}, \bibinfo{pages}{4149}
  (\bibinfo{year}{1998}).

\bibitem[{\citenamefont{Deutsch and Jessen}(1998)}]{jessen98_2}
\bibinfo{author}{\bibfnamefont{I.~H.} \bibnamefont{Deutsch}} \bibnamefont{and}
  \bibinfo{author}{\bibfnamefont{P.~S.} \bibnamefont{Jessen}},
  \bibinfo{journal}{Phys. Rev. A} \textbf{\bibinfo{volume}{57}},
  \bibinfo{pages}{1972 } (\bibinfo{year}{1998}).

\bibitem[{\citenamefont{Prudnikov et~al.}(2001)\citenamefont{Prudnikov,
  Taichenachev, Tumaikin, and Yudin}}]{pru2001}
\bibinfo{author}{\bibfnamefont{O.~N.} \bibnamefont{Prudnikov}},
  \bibinfo{author}{\bibfnamefont{A.~V.} \bibnamefont{Taichenachev}},
  \bibinfo{author}{\bibfnamefont{A.~M.} \bibnamefont{Tumaikin}},
  \bibnamefont{and} \bibinfo{author}{\bibfnamefont{V.~I.} \bibnamefont{Yudin}},
  \bibinfo{journal}{JETP} \textbf{\bibinfo{volume}{93}}, \bibinfo{pages}{63}
  (\bibinfo{year}{2001}).

\bibitem[{\citenamefont{Prudnikov
  et~al.}(1999{\natexlab{b}})\citenamefont{Prudnikov, Taichenachev, Tumaikin,
  and Yudin}}]{Prudnikov1999}
\bibinfo{author}{\bibfnamefont{O.~N.} \bibnamefont{Prudnikov}},
  \bibinfo{author}{\bibfnamefont{A.~V.} \bibnamefont{Taichenachev}},
  \bibinfo{author}{\bibfnamefont{A.~M.} \bibnamefont{Tumaikin}},
  \bibnamefont{and} \bibinfo{author}{\bibfnamefont{V.~I.} \bibnamefont{Yudin}},
  \bibinfo{journal}{JETP Letters} \textbf{\bibinfo{volume}{70}},
  \bibinfo{pages}{443} (\bibinfo{year}{1999}{\natexlab{b}}).

\bibitem[{\citenamefont{Kalganova et~al.}(2017)\citenamefont{Kalganova,
  Prudnikov, Vishnyakova, Golovizin, Tregubov, Sukachev, Khabarova, Sorokin,
  and Kolachevsky}}]{Kalganova}
\bibinfo{author}{\bibfnamefont{E.}~\bibnamefont{Kalganova}},
  \bibinfo{author}{\bibfnamefont{O.}~\bibnamefont{Prudnikov}},
  \bibinfo{author}{\bibfnamefont{G.}~\bibnamefont{Vishnyakova}},
  \bibinfo{author}{\bibfnamefont{A.}~\bibnamefont{Golovizin}},
  \bibinfo{author}{\bibfnamefont{D.}~\bibnamefont{Tregubov}},
  \bibinfo{author}{\bibfnamefont{D.}~\bibnamefont{Sukachev}},
  \bibinfo{author}{\bibfnamefont{K.}~\bibnamefont{Khabarova}},
  \bibinfo{author}{\bibfnamefont{V.}~\bibnamefont{Sorokin}}, \bibnamefont{and}
  \bibinfo{author}{\bibfnamefont{N.}~\bibnamefont{Kolachevsky}},
  \bibinfo{journal}{Phys. Rev. A} \textbf{\bibinfo{volume}{96}},
  \bibinfo{pages}{033418} (\bibinfo{year}{2017}).

\bibitem[{\citenamefont{Hemmerich and H\"ansch}(1993)}]{Hemmerich1993}
\bibinfo{author}{\bibfnamefont{A.}~\bibnamefont{Hemmerich}} \bibnamefont{and}
  \bibinfo{author}{\bibfnamefont{T.~W.} \bibnamefont{H\"ansch}},
  \bibinfo{journal}{Phys. Rev. Lett.} \textbf{\bibinfo{volume}{70}},
  \bibinfo{pages}{410} (\bibinfo{year}{1993}).

\bibitem[{\citenamefont{Monroe et~al.}(1990)\citenamefont{Monroe, Swann,
  Robinson, and Wieman}}]{Monroe1990}
\bibinfo{author}{\bibfnamefont{C.}~\bibnamefont{Monroe}},
  \bibinfo{author}{\bibfnamefont{W.}~\bibnamefont{Swann}},
  \bibinfo{author}{\bibfnamefont{H.}~\bibnamefont{Robinson}}, \bibnamefont{and}
  \bibinfo{author}{\bibfnamefont{C.}~\bibnamefont{Wieman}},
  \bibinfo{journal}{Phys. Rev. Lett.} \textbf{\bibinfo{volume}{65}},
  \bibinfo{pages}{1571} (\bibinfo{year}{1990}).

\bibitem[{\citenamefont{Gibble et~al.}(1992)\citenamefont{Gibble, Kasapi, and
  Chu}}]{Chu1992}
\bibinfo{author}{\bibfnamefont{K.~E.} \bibnamefont{Gibble}},
  \bibinfo{author}{\bibfnamefont{S.}~\bibnamefont{Kasapi}}, \bibnamefont{and}
  \bibinfo{author}{\bibfnamefont{S.}~\bibnamefont{Chu}}, \bibinfo{journal}{Opt.
  Lett.} \textbf{\bibinfo{volume}{17}}, \bibinfo{pages}{526}
  (\bibinfo{year}{1992}).

\bibitem[{\citenamefont{Javanainen}(1992)}]{Langevin}
\bibinfo{author}{\bibfnamefont{J.}~\bibnamefont{Javanainen}},
  \bibinfo{journal}{Phys. Rev. A} \textbf{\bibinfo{volume}{46}},
  \bibinfo{pages}{5819} (\bibinfo{year}{1992}),
  \urlprefix\url{https://link.aps.org/doi/10.1103/PhysRevA.46.5819}.

\bibitem[{\citenamefont{Prudnikov and
  Arimondo}(2004{\natexlab{b}})}]{Langevin2}
\bibinfo{author}{\bibfnamefont{O.~N.} \bibnamefont{Prudnikov}}
  \bibnamefont{and} \bibinfo{author}{\bibfnamefont{E.}~\bibnamefont{Arimondo}},
  \bibinfo{journal}{J. Opt. B: Quantum Semiclass. Opt.}
  \textbf{\bibinfo{volume}{6}}, \bibinfo{pages}{336}
  (\bibinfo{year}{2004}{\natexlab{b}}).

\bibitem[{\citenamefont{Schünemann et~al.}(1998)\citenamefont{Schünemann,
  Engler, Zielonkowski, Weidemüller, and Grimm}}]{SCHUNEMANN1998263}
\bibinfo{author}{\bibfnamefont{U.}~\bibnamefont{Schünemann}},
  \bibinfo{author}{\bibfnamefont{H.}~\bibnamefont{Engler}},
  \bibinfo{author}{\bibfnamefont{M.}~\bibnamefont{Zielonkowski}},
  \bibinfo{author}{\bibfnamefont{M.}~\bibnamefont{Weidemüller}},
  \bibnamefont{and} \bibinfo{author}{\bibfnamefont{R.}~\bibnamefont{Grimm}},
  \bibinfo{journal}{Optics Communications} \textbf{\bibinfo{volume}{158}},
  \bibinfo{pages}{263} (\bibinfo{year}{1998}), ISSN \bibinfo{issn}{0030-4018}.

\bibitem[{\citenamefont{Hulet et~al.}(2020)\citenamefont{Hulet, Nguyen, and
  Senaratne}}]{Randall2020}
\bibinfo{author}{\bibfnamefont{R.}~\bibnamefont{Hulet}},
  \bibinfo{author}{\bibfnamefont{J.~H.~V.} \bibnamefont{Nguyen}},
  \bibnamefont{and}
  \bibinfo{author}{\bibfnamefont{R.}~\bibnamefont{Senaratne}},
  \bibinfo{journal}{Review of Scientific Instruments}
  \textbf{\bibinfo{volume}{91}}, \bibinfo{pages}{011101}
  (\bibinfo{year}{2020}).

\bibitem[{\citenamefont{Hernandez-Rajkov
  et~al.}(2020)\citenamefont{Hernandez-Rajkov, Padilla-Castillo, Mendoza-Lopez,
  Colin-Rodriguez, Gutierrez-Valdes, Morales-Ramirez, Gutierrez-Arenas,
  Gardea-Flores, Jauregui-Renaud, Seman et~al.}}]{Rajkov2020}
\bibinfo{author}{\bibfnamefont{D.}~\bibnamefont{Hernandez-Rajkov}},
  \bibinfo{author}{\bibfnamefont{J.~E.} \bibnamefont{Padilla-Castillo}},
  \bibinfo{author}{\bibfnamefont{M.}~\bibnamefont{Mendoza-Lopez}},
  \bibinfo{author}{\bibfnamefont{R.}~\bibnamefont{Colin-Rodriguez}},
  \bibinfo{author}{\bibfnamefont{A.}~\bibnamefont{Gutierrez-Valdes}},
  \bibinfo{author}{\bibfnamefont{S.~A.} \bibnamefont{Morales-Ramirez}},
  \bibinfo{author}{\bibfnamefont{R.~A.} \bibnamefont{Gutierrez-Arenas}},
  \bibinfo{author}{\bibfnamefont{C.~A.} \bibnamefont{Gardea-Flores}},
  \bibinfo{author}{\bibfnamefont{R.}~\bibnamefont{Jauregui-Renaud}},
  \bibinfo{author}{\bibfnamefont{J.~A.} \bibnamefont{Seman}},
  \bibnamefont{et~al.}, \bibinfo{journal}{Revista mexicana de física}
  \textbf{\bibinfo{volume}{66}}, \bibinfo{pages}{388} (\bibinfo{year}{2020}).

\end{thebibliography}

\end{document}